\documentclass[aps,physrev,graphicx,amsmath,amssymb,reprint]{revtex4-2} % Two-column 'reprint'
\usepackage{graphicx}
\usepackage{mathtools}

\usepackage[version=3]{mhchem} % Formula subscripts using \ce{}
\usepackage{dcolumn}% Align table columns on decimal point
\newcolumntype{d}[1]{D{.}{\cdot}{#1} }
\usepackage{bm}% bold math
\usepackage{xcolor,soul}
\usepackage{siunitx}
\usepackage{float}
\usepackage{physics}
\usepackage[font=small,labelfont=bf, justification=Justified, singlelinecheck=off, format=plain]{caption} % 'format=plain' avoids hanging indentation
\usepackage{subcaption} 
\usepackage{tikz}
\usetikzlibrary{decorations.markings}

\usepackage{mathpazo} %JMF - Palatino + math font :^)

\usepackage{siunitx}% JMF addition - once a physicist, always a physicist
\usepackage[parfill]{parskip}
\parskip = \baselineskip

\usepackage{scalerel}

\newcommand*{\paral}{\stretchrel*{\parallel}{\perp}}

\begin{document}
\title{
Predicting polaron mobility in organic semiconductors with the Feynman variational approach
%Working title: Project Ruby Polarons \\
%Working title: Predicting organic semiconductor charge-carrier mobility with the Feynman polaron approach
%Working title: How I learnt to stop worrying and love the Feynman variational approach
}
\date{\today} % useful to track hard copy of drafts

\author{Bradley A. A. Martin}
\author{Jarvist Moore Frost}

\affiliation{Department of Physics, Imperial College London, Exhibition Road, London  SW7 2AZ, UK}
\affiliation{Department of Chemistry, Imperial College London, Exhibition Road, London SW7 2AZ, UK}
\email[Electronic mail:]{jarvist.frost@imperial.ac.uk}


\begin{abstract}
% Organic polarons woo woo

We extend the Feynman variational method applied to the parabolic-band Fr\"ohlich (continuum) large polaron~\cite{Feynman1955} to a Holstein (lattice) small polaron, with a parabolic-band. This new theory shows a discrete localisation as a function of coupling strength. Having build the theory with the same quasi-particle Lagrangian as the 1955 work, we can directly use the FHIP~\cite{Feynman1962} response theory to calculate DC mobility and complex conductivity. We show that we can take matrix elements from electronic structure calculations on real materials, by modelling charge-carrier mobility in crystalline Rubrene. Good agreement is found to measurement, with a predicted mobility of $\mu = 47.72$~\si{cm^2 V^{-1} s^{-1}} at $300$~\si{K}.  

\end{abstract}

% 71.38.-k	Polarons and electron-phonon interactions
% 71.20.Nr 	Semiconductor compounds 
% 71.38.Fp	Large or Fröhlich polarons
% 63.20.Kr	Phonon-electron and phonon-phonon interactions

% 72.40.+w 	Photoconduction and photovoltaic effects
% 61.66.Fn 	Inorganic compounds 
% 88.40.jn 	Thin film Cu-based I-III-VI2 solar cells
% 88.40.-j 	Solar energy

\maketitle 

\section{Introduction}\label{introduction}

Polarons are the quasi-particles formed by an electron (or hole) interacting with a polar material. The distortion of the material (driven by the electron-phonon coupling) tries to localise the charge carrier with the back-reaction. 
If the electron-phonon coupling is relatively weak, the polaron is adiabatically connected to the free carrier Bloch state. Perturbation theory built on this delocalised state can be used to evaluate mobility. 
As momentum is a well-defined quantum number, charge transport is coherent, and charge carrier mobility reduces with temperature, due to increased scattering.
For strong electron-phonon coupling, the polaron is localised to a small region. Charge transport is incoherent, and occurs through temperature-activated hopping. 

In organic electronic materials, the electron-phonon coupling is relatively
strong versus the small kinetic energy of the charge carrier (i.e. a large effective mass, or equivalently a small transfer integral). Generally, this gives rise to a small-polaron, driven by local (rather than long-range dielectric) electron-phonon coupling. 

Recently the transient localisation theory~\cite{Fratini2016} combines band and hopping transport models via dynamic disorder, where thermal fluctuation leads to transiently localised states. An approach is developed in the relaxation time approximation, where propagation between these localisation events is coherent. 

To bypass the empirical or modelling assumptions of these approaches, solving the non-adiabatic dynamics directly is necessary. This becomes a full quantum-field problem as you are simulating a single electron in the simplest case, but interacting with an infinite field of lattice vibrations (phonon quantisations). 

% JMF: Text not ready enough for v1
%Limiting the simulation of non-adiabatic dynamics to surface-hopping\cite{Tulley}, where a classical molecular dynamics trajectory is followed, with forces from a particular excited electronic state surface, and a stochastic hopping between electronically excited states to reproduce the correct partition function (??? succinct? correct?) ...
%Using this method, and simulating the spread of the non-adiabatic wavefunction in a sufficiently large volume, a mobility can be inferred by the Einstein relation. 
%These simulations have quite significant researcher and computer time in their setup and execution. 

% Samuelle Paper c.f. ; rubrene

% Yang 2018 paper - Rubrene, Pentancene, C60 :^)

Recently Giannini and coworkers~\cite{Giannini2019,Giannini2020} have developed a non-adiabatic surface-hopping-based method, combined with electronic structure evaluation of the adiabatic and non-adiabatic electron transfer integrals, to predict mobility in crystalline semiconductors. 
A limitation of the method is that the simulation box must be sufficient to fully enclose the charge-carrier state~\cite{Giannini2020}, and then the mobility is inferred by diffusion. 

%Other approaches include .... <literature dump>

Chang et al.~\cite{Chang2022} recently developed a cumulant expansion of many-body-perturbation-theory combined with a linear response mobility method, to estimate the temperature-dependent electron mobility in crystalline Naphthalene. 

One polaron theory that applies from weak to strong coupling is the celebrated Feynman variational approximation~\cite{Feynman1955}. 
Formulated in the model Fr\"ohlich Hamiltonian for a continuum polaron with
dielectrically mediated electron-phonon coupling (i.e. the infrared activity of phonon modes), this variational theory is accurate to a few percent in energy from weak to strong coupling~\cite{Hahn2018}. The method is based on the path-integral formulation of quantum mechanics, where the infinite degrees of freedom of the phonon field are integrated out. The Lagrangian action associated with the true Fr\"ohlich Hamiltonian cannot be analytically integrated over, so instead a variational quasi-particle solution is made of the electron (hole) attached by a spring to a fictitious mass. This fictitious mass represents the phonon drag term. 

FHIP~\cite{Feynman1962} builds a linear-response theory (relatively low field) around this variational solution of the polaron quasi-particle, for the Fr\"ohlich Hamiltonian. The original electron-phonon Hamiltonian is not directly present, but corrections to the response function are made by expanding the true influence functional about the influence functional of the trial quasi-particle. This quasi-particle solution is parameterised by two variational parameters $v$ and $w$ (or equivalently, the fictitious mass $M$ and spring coupling constant $C$). The interacting thermal bath of \emph{harmonic} phonon excitations is replaced with a two-point memory function. Thereby a mobility theory is developed which contains all orders of
perturbation theory, and which can be evaluated extremely efficiently. 

In crystalline semiconductors, these techniques are being increasingly
recognised as powerful and predictive, not least as polar optical mode scattering (i.e. polaronic effects) is recognised as a major scattering process across a wide range of technologically relevant materials. 

FHIP works equally well~\cite{Hellwarth1999,Frost2017} with the finite temperature \=Osaka~\cite{Osaka1959} extension to Fr\"ohlich's Hamiltonian, even though the parameters of the variational solution are observed to be quite different~\cite{Frost2017}. This begs the question of whether the FHIP approach can be used with other quasi-particle mobility problems, and whether the original Feynman variational solution can be extended beyond Fr\"ohlich's Hamiltonian to other electron-phonon couplings. 

We thought it would be a good idea to see whether we could apply these polaron variation theories to organic electronic materials, where the polaron problem is treated on a lattice of individual sites (rather than a continuum), with Holstein (on site) and Peierls (off-site) electron-phonon couplings, rather than the dielectric electron-phonon coupling in the Fr\"ohlich Hamiltonian (the physics of a polar continuum). The simplest form of such a lattice polaron problem is the Holstein Hamiltonian~\cite{HolsteinI1959, HolsteinII1959}.

To demonstrate our new theory, we make predictions of the behaviour of Rubrene, a high-mobility organic crystalline semiconductor. We take the coupling matrix elements from a recent electronic structure calculation~\cite{Ordejn2017}. Rubrene has been relatively well studied, and
high purity macroscopic single crystals can now be prepared, indicating that
our predictions of finite-temperature finite-frequency mobility and optical
absorption will be testable soon. 

\section{Polaron Theories}

Kornilovitch and coworkers studied the lattice polaron in both the Holstein-Peierls~\cite{kornilovitch_polaron_1997} and 
Fr\"ohlich~\cite{alexandrov_mobile_1999} form. Their key approach~\cite{kornilovitch_path_2007} is to use the 1955 Feynman
trick~\cite{Feynman1955} of integrating over the phonon field but
evaluate the path-integral construction of the partition function numerically
with Monte-Carlo. %\textcolor{red}{JMF: (??? to be fleshed out)}

Coropceanu et al.~\cite{Coropceanu2007} review the standard approaches to calculating charge transfer parameters in organic semiconductors. Most organic semiconductors are highly disordered or amorphous, so the first step to calculate charge carrier mobility is to evaluate the transfer of an electron between two discrete molecules in a vacuum. For an organic crystal, this naturally maps to a tight-binding model with a discrete site-basis of the individual molecules.

\subsection{General Polaron}

The Hamiltonian for a general (single conduction or valence band) polaron model~\cite{Alexandrov2009} can be
written in momentum-basis and second-quantised form as 
\begin{equation}
    \begin{aligned}
        H &= \sum_{\vb{k}} \epsilon_{\vb{k}} c^\dagger_{\vb{k}} c_{\vb{k}}
        + \sum_{\vb{q}} \hbar \omega_{\vb{q}} b^\dagger_{\vb{q}} b_{\vb{q}} \\
        &+ \sum_{\vb{k},\vb{q}} V_{\vb{k}, \vb{q}} c^\dagger_{\vb{k}+\vb{q}}
        c_{\vb{k}} (b^\dagger_{\vb{-q}} + b_{\vb{q}}) ,
    \end{aligned}
\end{equation} 
where $\epsilon_{\vb{k}}$ is the electron band energy for momentum $\vb{k}$,
$c^\dagger_{\vb{k}}$ and $c_{\vb{k}}$ are the electron creation and
annihilation operators for an electron with momentum $\vb{k}$,
$\omega_{\vb{q}}$ is the phonon frequency for momentum $\vb{q}$,
$b^\dagger_{\vb{q}}$ and $b_{\vb{q}}$ are the phonon creation and annihilation operators for a phonon with momentum $\vb{q}$, $V_{\vb{k}, \vb{q}}$ is the electron-phonon coupling matrix which describes the strength of the interaction.

To apply the 1955 Feynman variational method~\cite{Feynman1955}, we must simplify this Hamiltonian to a single effective-mass electron, interacting with a dispersion-less optical phonon of frequency $\omega_0$, 
\begin{equation}
    H = \frac{\vb{p}^2_{el}}{2m_b} + \hbar\omega_{0} \sum_{\vb{q}} 
    b^{\dagger}_{\vb{q}}b_{\vb{q}} + \sum_{\vb{q}} V_{\vb{q}} \rho_{\vb{q}} \left( b^\dagger_{-\vb{q}} + b_{\vb{q}} \right). 
\end{equation}
Here $\vb{p}_{el}$ is the momentum operator of the electron/hole, 
$m_b$ is the bare-band effective mass. 
The bare-band effective masses are typically obtained by a parabolic fit to the
band-structure of an electronic structure calculation.
The density operator, $\rho_{\vb{q}}$, is $\rho_{\vb{q}} = \sum_{\vb{k}}
c^\dagger_{\vb{k}+\vb{q}} c_{\vb{k}}$ in second-quantisation, and
$\rho_{\vb{q}} = e^{i \vb{q} \cdot \vb{r}_{el}}$ in first-quantisation.

\subsection{Large Polarons: The Fr\"ohlich Model}

The Fr\"ohlich model~\cite{frohlich} for large polarons approximates the lattice as a continuum, with continuous and unbounded electron and phonon wave vectors, $-\infty < k, q < \infty$. The electron dispersion is parabolic with an effective mass $m_F$,
\begin{equation}
    \epsilon_k^{(F)} = \frac{\hbar^2 k^2}{2 m_F} = \frac{p^2_{el}}{2 m_F}.
\end{equation}
The electron-phonon coupling is the long-range dielectric response of the
lattice modelled as a continuum dielectric. In $n$-dimensions this is~\cite{Peeters1986}
\begin{equation}
    \begin{aligned}
        V^{(F)}_{\vb{q}} &= \frac{g_{F}}{\sqrt{V q^{n-1}}}, \\
        &= \hbar \omega_0 \left(\frac{\alpha_F r_p}{V} \Gamma\left(\frac{n-1}{2}\right) \left( \frac{2 \sqrt{\pi}}{q} \right)^{n-1}\right)^{\frac{1}{2}},
    \end{aligned}
\end{equation}
where $g_{F}(n)$ is the Fr\"ohlich electron-phonon coupling strength, and $V$
the crystal volume in $n$ dimensions. 
The unitless Fr\"ohlich electron-phonon coupling constant  $\alpha_F$ can be
calculated from bulk material properties using 
\begin{equation}
    \alpha_F = \frac{1}{2} \frac{1}{4\pi \varepsilon_{\text{vac}}} 
    \left(\frac{1}{\varepsilon_{\infty}} - \frac{1}{\varepsilon_0}\right)
    \frac{e^2}{\hbar \omega_0 r_p},
\end{equation}
where $e$ is the electron charge, $\varepsilon_{\text{vac}}$, $\varepsilon_{\infty}$ and $\varepsilon_0$ are the vacuum, optical and static dielectric constants, and $r_p = \sqrt{\hbar / 2 m_b \omega_0}$ is the characteristic polaron length.

\subsection{Small Polarons: The Holstein Model}

The Holstein model~\cite{HolsteinI1959, HolsteinII1959} is on a lattice, with
$N$ discrete electron and phonon wave vectors, confined within the first
Brillouin zone $-\frac{\pi}{a} < k, q < \frac{\pi}{a}$. 
$N$ is the number of sites and $a$ the lattice constant. 
Following a tight-binding description, the electron dispersion relation is,
\begin{equation}
    \epsilon^{(H)}_{\vb{k}} = -2 J \sum_{i=1}^n \cos(k_i) ,
\end{equation}
where $J$ is the nearest-neighbour electronic hopping amplitude. 
The electron-phonon coupling $V_{\vb{q}} \to V_0$ is a constant,
\begin{equation}
    V^{(H)}_{0} = \frac{g_{H}}{\sqrt{N}} = \sqrt{\frac{2 n J \hbar \omega_0 \alpha_H}{N}} ,
\end{equation}
where the Holstein electron-phonon coupling energy $g_{H}$ is parameterised by the \emph{unitless} coupling constant $\alpha_H = g_H^2 / 2 n J \hbar \omega_0$. 
As both $J$ and $\omega_0$ are natural energy scales, we define
a second unitless coupling parameter, the `adiabaticity', as $\gamma \equiv \hbar \omega_0 / J$. Since we have an explicit expression for the effective band mass, we can also express the characteristic polaron length as $r_p = a / \sqrt{\gamma}$.

\subsection{Marcus theory of small polaron hopping}

Marcus theory~\cite{Marcus1956} is the standard approach to model small-polaron hopping transport in organic semiconductors~\cite{Nelson2009}. 
The theory models the high-temperature limit of thermally activated hopping. 
This reorganisation energy can be understood as the electron-phonon coupling. The reorganisation energy is split into, using the terms of Marcus' theory developed for solvated reactions, inner and outer sphere reorganisation energies. Jortner~\cite{Jortner1976} shows how the semi-classical Marcus theory is consistent with the low-temperature quantum mechanical tunnelling limit. 
%\textcolor{red}{(JMF: These are just stubs, to cite the papers ahead of properly reading + explaining).}

The inputs to the theory, which can be calculated with electronic structure techniques, are the transfer integral (kinetic energy) between the electron localised on the two states, and a reorganisation energy. 

The inner sphere reorganisation energy, for a molecular semiconductor, is driven by the change in bond lengths of the molecule upon charging and discharging. This is equivalent to the Holstein (same site) electron-phonon coupling. 

The outer-sphere reorganisation energy contains contributions from vibration-driven fluctuation in the transfer integral and is equivalent to the Peierls (off-site) electron-phonon coupling.

%\subsection{Calculation of band structure}

%The electronic coupling between organic electronic materials is usually described in terms of a transfer integral (hopping matrix element) between two adiabatic states localised on nearest-neighbour molecules. 
%Polaron theories based in a band description instead consider a band-structure of Bloch states as their starting point, with an associated effective mass that describes the quadratic dispersion relationship observed near the extremal points. 

%To link these concepts, we use a simple tight-binding model of the electronic structure. 
%In one dimension, the coupling $t$ between successive lattice locations (Rubrene molecules in our example) gives rise to a dispersion relation (band structure),
%\begin{equation}
%E(\vb{k}; J, a) = \epsilon_0 - 2 z J \sum_{x=1}^n \cos(k_{x} a) .   
%\end{equation}
%where $z$ is the number of nearest neighbouring lattice sites and $a$ is the lattice constant. Expanding around $k=0$ for a given orthogonal direction in k-space and using the identity $\cos(\theta) \approx 1 - \frac{\theta^2}{2} \quad ; \quad \theta \ll 1$, then by analogy with the free electron dispersion relationship $E(k)=\frac{p^2}{2m_e}=\frac{\hbar^2k^2}{2m_e}$ we have the tight-binding effective mass as, 
%\begin{equation} \label{eqn:effmass}
%    m_b = \frac{\hbar^2}{2 J a^2} .
%\end{equation}

\subsection{Calculation of electron-phonon couplings}

% Should probably have something about the methods of Ordejon in here
% Or at least some useful description of how the \hbar\omega g factors sum up to a Lambda ?

The reorganisation energy, often calculated with the 'four-point' method by considering the relaxation of a molecule in its charged and uncharged states, can also be composed of the sum of individual phonon modes and their dimensionless electron-phonon coupling, 
\begin{equation}
\lambda = \sum_i^N \hbar \omega_i \alpha_i ,
\end{equation}
where the sum is over N phonon (vibrational) modes, $\hbar$ the reduced Planck constant, $\omega_i$ the reduced frequency of the $i$'th mode, $\alpha_i$ the dimensionless electron-phonon coupling of the $i$'th mode. 

Additionally, this can be split into an `inner sphere', and `outer sphere' reorganisation energy and are equivalent to the Holstein (on-site/intramolecular) and Peierls (off-site/intermolecular) electron-phonon couplings respectively.

Marcus theory makes an adiabatic approximation and assumes that the rate of electron transfer is slow relative to the vibrational $w_i$ degrees of freedom, so this information is not required. The Feynman theory explicitly considers the exchange of energy between the electron and the vibrational degrees of freedom so we must retain information about $w_i$, at least in some `effective frequency' approach as done by Hellwarth et al.~\cite{Hellwarth1999}.

% \subsection{The Fr\"ohlich Hamiltonian for organic electronic materials}

\section{A Path Integral for Small Polarons}

In the 1980s, Hans De Raedt and Ad Lagendijk~\cite{Raedt1983, Raedt1985}
derived the discrete-time path integral for the small lattice polaron, which
was then further developed in the 1990s by Pavel Kornilovitch~\cite{kornilovitch_polaron_1997,
kornilovitch_continuous-time_1998, kornilovitch_ground-state_1999,
kornilovitch_giant_1999, kornilovitch_band_2000, kornilovitch_mass_2004,
kornilovitch_path_2007}. 
Kornilovitch derived the continuous path integral limit of the lattice polaron
and developed a Continuous-time Path Integral Monte Carlo method for
calculating properties of small polarons, with a special focus on the Holstein
model and a lattice version of the Fr\"ohlich model (which allows for
long-range electron-phonon coupling). 

The path integral for the partition function of a system is a sum of all possible position and momentum paths in phase space. In the continuum large-polaron model, the sum over all momentum paths can be made as the momentum paths are continuous and unbounded, and the parabolic electronic dispersion is quadratic in momentum. Therefore, the corresponding momentum path integral is an evaluable Gaussian functional integral.

For the small lattice polaron, we have two main difficulties. First, the electronic dispersion is that of a tight-binding model and is not quadratic.  Second, the momentum and position paths are discrete and the `path integral' is an infinite summation rather than a functional integral. Therefore, the corresponding `path integral' is a non-Gaussian summation rather than a typical Gaussian functional integral. To apply the Feynman variational approximation, the trial path integral must have the same measure as the original system, but since the paths are discrete, even if the trial path-integral were Gaussian, a Gaussian sum does not have a closed-form expression, unlike a Gaussian integral. Even if we approximate the paths to be continuous to bypass this issue, unlike for the Fr\"ohlich model, we cannot choose the kinetic action of the trial path-integral to be identical to the Holstein kinetic action as it is non-Gaussian. We would then have an additional term, the difference between the trial and true kinetic actions. To apply the Feynman-Jensen inequality, we need this term to be convex; this is no longer guranteed.

In the following section, we present the derivation of the discrete-time path integral for the Holstein model to demonstrate these issues. We then take the continuous-time limit and show that the continuous-time path-integral has the same form as the standard phase-space path-integral but with the Holstein Hamiltonian. Finally, we make an effective-mass approximation for the Holstein kinetic action so that the Feynman-Jensen variational approach can be used.

\subsection{Discrete-Time Holstein Path Integral}

% To use the Feynman-Jensen variational approximation and in analogy with the
% Fr\"ohlich model, we make an effective-mass approximation. 
% A Taylor expansion of the tight-binding band structure and using the identity
% $\cos(\theta) \approx 1 - \theta^2 / 2 $ for $\theta \ll 1$, gives the parabolic electronic dispersion $\epsilon_k = p^2 / 2 m_e = \hbar^2 k^2 / 2 m_e$. 
% Our tight-binding effective band-mass for the Holstein model $m_H$ is 
% \begin{equation}
%     m_H = \frac{\hbar^2}{2 J a^2} ,
% \end{equation}
% with a dispersion
% \begin{equation}
%     \epsilon^{(H)}_{\vb{k}} \approx \frac{p^2_{el}}{2 m_H} .
% \end{equation}
We begin with the Holstein Hamiltonian in a mixed representation, 
\begin{equation}
    \begin{aligned}
        H &= H_{ph} + H_{el-ph} + H_{el} , \\
        H_{ph} &= \frac{1}{2m_{ph}} \sum_{i=1}^N p_i^2 + \frac{m_{ph} \omega_0^2}{2} \sum_{i=1}^N x_i^2,  \\
        H_{el-ph} &= g_H \sum_{n=1}^N x_i c^\dagger_i c_i , \\
        H_{el} &= -J \sum_{i=1}^N c^\dagger_i c_{i+1} + c_{i+1}^\dagger c_i ,
    \end{aligned}
\end{equation}
where the phonons are expressed in terms of their momenta $p_i$ and positions
$x_i$ where $i$ labels the corresponding lattice site. $m_{ph}$ is the mass of one
lattice-site (here we assume all of them to have the same mass) and $\omega_0$
is the dispersionless phonon frequency where we assume to have only one mode
(i.e. an Einstein mode). 
The electron-phonon coupling is $g_H$ and is linear in the lattice site
distortion ($x_i$) and acts entirely locally. 
The electron description remains in terms of the creation and annihilation operators $c^\dagger_i$, $c_i$ on a lattice-site $i$.

In the derivation of the path integral, the quantum statistical partition
function for a system may be obtained by inserting successive resolutions of
the identity within the definition of a quantum trace. In the limit of an
infinite number of insertions ($M\to\infty$), the Trotter-Suzuki
expression~\cite{Trotter1959, Hatano2005} gives a direct equality between this
discretised partition function $Z_M$ and the full partition function $Z$. 
\begin{equation}
    \begin{aligned}
        Z &\equiv \Tr{e^{-\beta H}} = \lim_{M\to\infty} Z_M , \\
        Z_M &= \Tr{\left[ e^{-\Delta\tau H_0} e^{-\Delta\tau H_1} e^{\Delta\tau H_2} \right]^M} ,
    \end{aligned}
\end{equation}
where $M$ is the number of time-slices and $\Delta\tau = \beta / M$ is the imaginary-time between time-slices. 

For the case of a lattice polaron, we have the time-discretised partition function,
\begin{equation}
    Z_M = c_1 {\Delta\tau}^{-\frac{MN}{2}} \sum_{\{r_j\}} \int \left\{ \prod_{j=1}^M \prod_{n=1}^N dx_{n,j} \right\} e^{S_{ph}} \prod_{l=1}^M I_{\Delta r_l}(2 \tau J) ,
\end{equation}
where $\Delta r_l = r_{l+1} - r_l$ is the change in the electron position across one time-slice and is an integer value (i.e. multiple of the lattice constant).

The discretised Boson action is 
\begin{equation}
    S_{ph} = \sum_{j=1}^M \sum_{n=1}^N \left( \frac{\left( \Delta x_{n,j} \right)^2}{2 \Delta\tau} + \frac{\Delta\tau \omega_0^2 x^2_{n,j}}{2} + \Delta\tau x_{n,j} \delta_{n,r_j/a}\right) .
\end{equation}

The kinetic portion of the discretised action for the Fermion on a lattice is 
\begin{equation}
    I_{\Delta r_l}(2 \tau J) = \frac{1}{N} \sum_{n=1}^N \cos\left( \frac{2\pi n \Delta r_l}{N a}  \right) \exp\left(-z \cos\left(\frac{2\pi n}{N}\right)\right) ,
\end{equation}
where $a$ is the lattice constant. This function is a discrete form of the modified Bessel function of the first-kind $I_m(z)$~\cite{NIST:DLMF} where here we have $m = \Delta r_l = r_{l+1} - r_l$ and $z = 2 \tau J$. In the thermodynamic limit, $N \to \infty$, this becomes exactly the normal modified Bessel function.

The Bosonic integrals are Gaussian, and so have closed-form solutions. By expanding the Bosonic coordinates in Fourier modes,
\begin{equation}
    x_{n,j} = \frac{1}{\sqrt{M}} \sum_{k=0}^{M-1} \nu_{n,j} \exp\left( \frac{2\pi j k}{M} \right) ,
\end{equation}
we can diagonalise the Bosonic action,
\begin{equation}
    \begin{aligned}
    S_{ph} &= \sum_{n=1}^N \sum_{k=0}^{M-1} \left( \frac{\abs{\nu_{n,j}}^2}{\Delta\tau D_k^{-1}} \right. \\
    +& \left. \frac{\Delta\tau g_H \nu_{n,k}}{\sqrt{M}} \sum_{j=1}^M \delta_{n,r_j / a} \exp\left(\frac{2\pi j k}{M}\right) \right) ,
    \end{aligned}
\end{equation}
where 
\begin{equation}
    D_k^{-1} = 1 - \cos\left(\frac{2\pi k}{M} \right) + \frac{{\Delta\tau}^2 \omega_0^2}{2} ,
\end{equation}
is the inverse of the free-phonon Green's function. Integrating over $\nu_{n,k}$ gives
\begin{equation}
    \begin{aligned}
        Z_M &= c_2 Z^{ph}_M Z_M^{el} , \\
        Z^{ph}_M &= \left( \prod_{k=0}^{M-1} D^{1/2}_k \right)^N , \\
        Z^{el}_M &= \sum_{\{r_j\}} \left( \prod_{j=1}^M I_{\Delta r_j}(2 \Delta\tau J) \right) \\
        &\times\exp \left( {\Delta\tau}^2 \sum_{i=1}^M \sum_{j=1}^M F(i - j) \delta_{r_i, r_j} \right) ,
    \end{aligned}
\end{equation}
where $c_2$ is just a collection of normalisation factors
which will drop out of any expectation values.

\begin{equation}
    F(l) = \frac{\Delta\tau g_H^2}{4M} \sum_{k=0}^{M-1} D_k \cos\left(\frac{2\pi k l }{M}\right) ,
\end{equation}
is the memory function that fully encodes the electron-lattice interaction over all imaginary times.

Now the kinetic portion of the action is 
\begin{equation}
    K_M[r_j] = \sum_{j=1}^M \ln\left\{I_{\Delta r_j}(2J\Delta\tau)\right\}.
\end{equation}

Even for a free electron, this path integral does not have a closed-form solution. Regardless, we could choose the kinetic part of the trial action to be quadratic still, with an effective mass $m_e$, 
\begin{equation}
    K_{M,\text{trial}}[r_j] = \frac{m_e}{2} \sum_{j=1}^M \left(\frac{\Delta
    r_j}{\Delta\tau}\right)^2 \Delta \tau .
\end{equation}
The quasi-particle mass term $m_e$ is a variational parameter, to minimise the difference
between the trial and model actions. However, as the electron positions are discrete (restricted to the lattice positions), the trial path integral would be a discrete Gaussian sum, not a Gaussian integral:
\begin{equation}
    Z_{M,\text{trial}} \sim \sum_{\{r_j\}} \exp\left(K_{M,\text{trial}}[r_j]\right).
\end{equation}

This sum has no closed-form solution and would have to be evaluated
numerically, defeating our goal of using the Feynman variational method to
leverage computationally efficient analytic solutions, and directly apply the
FHIP theory. 

Therefore, since we aim to use the Feynman variational method, we will take the continuum limit. First, we go to the continuous-time limit of the Holstein path integral.

\subsection{Continuous-Time Holstein Path Integral}

To obtain the continuous-time limit of the partition function, we first
explicitly take out the summation over $N$ lattice sites from the kinetic action. 
Doing so, we find that we have a product of these lattice-site summations for each time slice,
\begin{equation}
    \begin{aligned}
        Z^{el}_M &= \frac{1}{(2N)^M} \sum_{\left\{r_j\right\}} \sum_{\left\{n_j\right\}} \\
        &\exp\left\{i \sum_{j=1}^M \frac{2 \pi n_j}{N a} \Delta r_j \right. 
        - \left. z \sum_{j=1}^M \cos(\frac{2 \pi n_j}{N})\right\} ,
    \end{aligned}
\end{equation}
where $n_j$ now depends on $j$. 
We have used the fact that the kinetic action is even with respect to $n$, and
that $\cos(x) = (\exp(ix) + \exp(-ix))/2$, to change the limits of the $n_j$ summations from $\{n_j\ |\ n_j \in [1, N]\}$ to $\{n_j\ |\ n_j \in [-N,N],\ n_j \neq 0 \}$. Note that the summations over $n_j$ exclude $n_j = 0$. 

We now have a summation of all possible paths a particle can take on the
lattice within $M$ time-steps. From the kinetic action, we can see that $2\pi n_j / N a \equiv \Delta k_j$ plays the role of a discrete lattice-momenta multiplying the changes in the electron position $\Delta r_j$. 

Our electronic partition function is in the form of a discrete phase-space path integral. Therefore, in the continuous-time limit $M \to \infty$ and $\Delta \tau \to 0$, the partition function may be written
as 
\begin{equation}
    Z = \mathcal{C} Z_{ph} \sum_{r(\tau)} \sum_{k(\tau)} \exp{S_{\text{eff}}} ,
\end{equation}
where $\mathcal{C}$ is the accumulation of normalisation factors. The positions $r(\tau)$ and lattice-momenta $k(\tau)$ are still restricted to $N$ discrete values. The effective action is 
\begin{equation}
    \begin{aligned}
        S_{\text{eff}} &= K[k(\tau), r(\tau)] - V_{\text{eff}}[r(\tau)] , \\
        K &= -\frac{2J}{\hbar} \int_0^{\hbar\beta} d\tau\ \cos{\left(a k(\tau)\right)} + \frac{i}{\hbar} \int_0^{\hbar\beta} d\tau\ k(\tau) \Dot{r}(\tau) , \\
        V_{\text{eff}} &= \frac{g_H^2}{4 \hbar} \int_0^{\hbar\beta} \int_0^{\hbar\beta} d\tau d\tau' D_{\omega_0}(\tau - \tau') \delta_{r(\tau), r(\tau')} ,
    \end{aligned}
\end{equation}
where the summations over $j$ have become imaginary-time integrals and 
\begin{equation}
    \begin{aligned}
        \lim_{\Delta\tau \to 0} \left\{\Delta r_j\right\} \rightarrow & 
            r(\tau) , \\
        \lim_{\Delta\tau \to 0} \left\{\Delta r_j / \Delta \tau\right\} \rightarrow & 
            d r(\tau) / d\tau \equiv \Dot{r}(\tau), \\
        \lim_{\Delta\tau \to 0} \left\{\Delta k_j \right\} \rightarrow
            & k(\tau) \equiv 2\pi n(\tau) / N a . 
    \end{aligned}
\end{equation}

Here $D_{\omega_0}(\tau)$ is the thermal (imaginary-time) phonon Green's function and is
\begin{equation} \label{eqn:phonongf}
    D_{\omega_0}(\tau) = \coth(\frac{\hbar\beta\omega_0}{2}) \cosh(\omega_0 \tau) - \sinh(\omega_0\tau).
\end{equation}

So far, we have derived the discrete-time Holstein path integral by starting with the Holstein Hamiltonian in a mixed representation, including phonon, electron-phonon coupling, and electronic components. We used the Trotter-Suzuki expression to discretise the quantum statistical partition function into a form suitable for path integral methods. We then expressed the partition function in terms of Bosonic and Fermionic actions, applied Gaussian integration to the Bosonic coordinates, and incorporated a variational trial action for the kinetic part of the path integral. Finally, we transitioned towards the continuum limit by reformulating the partition function in terms of continuous imaginary-time integrals, leading to an effective action that encapsulates the interactions and dynamics of the electron-phonon system.

\subsection{Thermodynamic and Continuum Limits}

In the thermodynamic limit $N\to\infty$, the summation over all discrete $k$ paths
becomes continuous and can be identified with the electron quasi-momentum
$k(\tau)$. The discrete sums over $k$ become continuous functional (path)
integrals over quasi-momentum paths confined to the first Brillouin Zone.
\begin{equation}
    Z = \mathcal{C} Z_B \int_{r \in a \mathbf{Z}} \mathcal{D}r(\tau) \int_{k \in [-\frac{\pi}{a},\frac{\pi}{a}]}
    \mathcal{D} k(\tau)\ e^{S_{\text{eff}}} .
\end{equation}
The electron position is still restricted to the lattice, represented above by
the set of integers, $\mathbf{Z}$, multiplied by the lattice constant, $a$. 

In the continuum limit $a \to 0$, the electron position can take any real value $r \in \mathbf{R}$ and the momentum is unbounded, $k \in (-\infty, \infty)$. Here we assume that $a \ll 1$ so that these conditions on the electron position and momentum are approximately satisfied.

This partition function is the standard representation of the phase-space path integral, 
\begin{equation}
    \begin{aligned}
        Z = \mathcal{C} \int \mathcal{D}r(\tau) \int &\mathcal{D}k(\tau) \exp\left\{\frac{i}{\hbar} \int_0^{\hbar\beta} d\tau\ k(\tau) \Dot{r}(\tau) \right. \\
         &\left.- \frac{1}{\hbar}\int_0^{\hbar\beta} d\tau\ H_{\text{eff}}[r(\tau), k(\tau)] \right\} ,
    \end{aligned}
\end{equation}
where the Hamiltonian is the tight-binding Hamiltonian with an additional non-local effective interaction term,
\begin{equation}
    H_{\text{eff}} \approx 2J \cos(a k(\tau)) + \int_0^{\hbar\beta} d\tau'\
    D_{\omega_0} (\tau - \tau') \delta_{r(\tau),r(\tau')} .
\end{equation}

%\red{JMF: Quite an important point. Maybe more sign posting?}
We could have started with this phase-space path integral, substituted in the Holstein Hamiltonian and performed the path integration over the lattice coordinates to arrive at the same result. 
Therefore to generalise to higher dimensions we may substitute a higher-dimensional tight-binding Hamiltonian. 
The effective interaction term will be similar but with a generalised Kronecker-Delta dependent on vector positions $\Vec{r}(\tau)$.

\subsection{The Effective Mass (Parabolic-Band) Approximation}

We still face difficulty in applying the variational method. 
The presence of the cosine in the kinetic action (from the tight-binding band structure) renders the overall action non-convex, even in imaginary-time, so Jensen's inequality does not hold. 

To make progress, we assume that $a k \ll 1$ and thereby make a parabolic (effective-mass) approximation. 
Since we have made the continuum approximation, this is a reasonable assumption. From the cosine form of the tight-binding band structure, and using the
small-angle approximation $\cos(\theta) = 1-\theta^2$, 
\begin{equation}
    \cos(a k(\tau)) \approx 1 - \frac{[a k(\tau)]^2}{2} .
\end{equation}

In one dimension the kinetic action is approximated by,
\begin{equation}
    K = 2 J \hbar \beta - \frac{1}{2 m_b} \int_0^{\hbar\beta} d\tau \left[k(\tau)\right]^2 + i \int_0^{\hbar\beta} d\tau k(\tau) \Dot{r}(\tau) ,
\end{equation}

where the effective band-mass of the tight-binding Hamiltonian is $m_b = \hbar^2 / 2 J a^2$. 

By making this approximation, the functional integral over $k(\tau)$ is in the same Gaussian form as for a free particle and can be solved exactly. 

Overall, we get a parabolic-band Holstein effective action in $n$-dimensions as 
\begin{equation}
    \begin{aligned}
        S_{\text{eff}}^{(H)} &= \frac{m^{(H)}_b}{2} \int_0^{\hbar\beta} d\tau\ \vb{\Dot{r}}^2 \\
        &- \frac{g_H^2}{\hbar} \int_0^{\hbar\beta} \int_0^{\hbar\beta} d\tau d\tau' D_{\omega_0}(\tau - \tau') \delta_{\vb{r}(\tau), \vb{r}(\tau')}.
    \end{aligned}
\end{equation}
We exclude the $2nJ\hbar\beta$ term since this is just the band-minimum energy and is absorbed into the normalisation factor $\mathcal{N}$ for the partition function. 

We reiterate the approximations we have made to derive this action, which we
will use in our numeric results. 
First, we went to the thermodynamic limit, which means we do not expect this model to capture any finite-size effects. 
Second, we make the continuum approximation so that the electron paths can be assumed to be continuous and the electron quasi-momenta unbounded. 
Third, we approximate the tight-binding (cosine) band with a parabolic band. 

This final effective-mass approximation means that there is a missing
contribution when we integrate across reciprocal space; we would expect it to
break down entirely neart the Brillouin-Zone boundaries. 
Nonetheless we have a truly short-range electron-phonon coupling.  
Upon integrating out the phonons, our short-range electron-phonon coupling
presents as a non-local point-like interaction of the electron with itself
through imaginary-time which is only non-zero when the electron crosses its
prior path.  

Since we have a Kronecker-delta like interaction, the \emph{phonon} momentum is
(correctly) bounded to remain within the first Brillouin zone. 
We can see this from the integral representation of the
Kronecker delta, 
\begin{equation}
    \delta _{r, r'} = \frac{a}{2\pi} \int_{-\pi/a}^{\pi/a} dq\ e^{i q (r - r')} .
\end{equation}
Therefore, as far as the phonons are concerned, we include a correct description of the lattice. 

We can generalise the Kronecker-delta to arbitrary dimensions $n$ in Cartesian coordinates 
\begin{equation}
    \delta_{\vb{r}, \vb{r'}} = \frac{V_n}{(2\pi)^n} \int_{-\pi/a}^{\pi/a} d\vb{q}\ e^{i \vb{q} \cdot (\vb{r} - \vb{r'})} ,
\end{equation}
where for example for a cubic unitcell $V_3 = a^3$. 

From this work on the Holstein model, we now have everything we need to
establish the machinery for a general variational method for polarons---one
that we can apply to an arbitrary electron-phonon Hamiltonian.

\subsection{General Polaron Path Integral}

With the machinary we have developed, we can now apply the Feynman variational
method to a polaron with a general electron-phonon interaction, provided we have a parabolic-band electron (or hole) linearly coupled to harmonic phonons. 
The path integral over the phonon operators is then Gaussian and can be evaluated analytically. The resultant electron action describes a temporally non-local self-interaction acting on the electron, 
\begin{equation} \label{eqn:eph-action}
    \begin{aligned}
        &S_{\text{pol}}[\vb{r}(\tau)] = \frac{m_b}{2} \int_0^{\hbar\beta} d\tau\ \Dot{\vb{r}}^2(\tau) \\
        &- \frac{1}{\hbar} \int_0^{\hbar\beta} d\tau \int_0^{\hbar\beta} d\tau'\ D_{\omega_0}(\tau - \tau') \Phi\left[\vb{r}(\tau), \vb{r}(\tau')\right] ,
    \end{aligned}
\end{equation}
where $ D_{\omega_0}(\tau)$ is the imaginary-time thermal phonon propagator and
the self-interaction functional is 
\begin{equation}
    \Phi\left[\vb{r}(\tau), \vb{r}(\tau')\right] = \sum_{\vb{q}} \abs{V_{\vb{q}}}^2 \rho_{\vb{q}} \left[\vb{r}(\tau)\right] \rho_{\vb{-q}}\left[\vb{r}(\tau')\right].
\end{equation}
Here $\rho_{\vb{q}}[\vb{r}(\tau)] = e^{i \vb{q} \cdot \vb{r}(\tau)}$ is the
density for the electron derived from corresponding first-quantisation density operator and $V_{\vb{q}}$ is a general electron-phonon coupling matrix element. The polaron self-interaction functional is where the specific electron-phonon coupling presents itself in our machinery. 

We now have a general polaron variation approach expression for the free energy, complex conductivity and DC mobility for arbitrary polaron models.  Before specialising we discuss the potential for direct numeric evaluation of the momentum integrals. 

Many closed-forms for the electron-phonon matrix $\abs{V_{\vb{q}}}^2$ and phonon dispersion $\omega_{\vb{q}}$ are known, such as for acoustic phonons, Bogoliubov-Fr\"hlich polaron, impurities etc. We could use such an analytic expression in our integrals, which may then admit closed-form solutions or be evaluated numerically. 

For real materials, we could instead use electronic structure methods to evaluate $\abs{V_{\vb{q}}}^2$ and $\omega_{\vb{q}}$ on a standard reciprocal space grid. These would then enter the variational method as arrays evaluated at the electron/hole band-extremum (e.g. the gamma-point $\vb{k} = \vb{0}$). Our $q$-integrands above would then become tensor products that are then concatenated over all $q$-points.

\subsection{Fr\"ohlich Polaron Path Integral}

As an example, we can express the Fr\"ohlich model in $n$ isotropic dimensions.
The self-interaction functional is 
\begin{equation}
    \begin{aligned}
        \Phi_F\left[\vb{r}(\tau), \vb{r}(\tau')\right] &= \sum_{\vb{q}} \frac{g_{F}^2}{V q^{n-1}} e^{i \vb{q} \cdot \left(\vb{r}(\tau) - \vb{r}(\tau') \right)} , \\
        &= g_{F}^2 \int \frac{d^n q}{(2\pi)^n} \frac{e^{i\vb{q}\cdot\left(\vb{r}(\tau) - \vb{r}(\tau')\right)}}{q^{n-1}} , \\
        &= \frac{g^2_F \abs{S^{n-1}}}{(2\pi)^{n}} \frac{1}{\abs{\vb{r}(\tau) - \vb{r}(\tau')}} ,
    \end{aligned}
\end{equation}
where $\abs{S^{n-1}} = 2\pi^{n/2}/\Gamma(n/2)$ is the hypervolume of the unit $(n-1)$-sphere and the phonon momentum is unbounded, $0 \leq q < \infty$. 
The Fr\"ohlich model makes the continuum approximation of the lattice for the phonon momentum, $\lim_{V \to \infty} V^{-1}\sum_{\vb{q}} \sim \int d^nq / (2\pi)^n$, where $V$ is the $n$-dimensional crystal volume.

\subsection{Holstein Polaron Path Integral}

For the main aim of this paper, the Holstein model in $n$ isotropic dimensions ($n$-dimensional hypercube) self-interaction functional is 
\begin{equation}
    \begin{aligned}
        \Phi_H\left[\vb{r}(\tau), \vb{r}(\tau')\right] &= g_H^2 \sum_{\vb{q}} e^{i \vb{q} \cdot \left(\vb{r}(\tau) - \vb{r}(\tau') \right)} ,\\
        &= g_{H}^2 a^n \int \frac{d^n q}{(2\pi)^n} e^{i\vb{q}\cdot\left(\vb{r}(\tau) - \vb{r}(\tau')\right)}, \\
        &= g_H^2 \delta^n_{\vb{r}(\tau)\vb{r}(\tau')} ,
    \end{aligned}
\end{equation}
where $\delta^n_{ij}$ is the $n$-dimensional Kronecker Delta functional and $-\pi/a
\leq q \leq \pi/a$.

\subsection{Summary}

We can now develop the variational path integral method for this generalised
polaron action and specialise to a specific electron-phonon Hamiltonian by using an explicit
expression for the electron-phonon coupling in the self-interaction functional
as we have above for the Fr\"ohlich and Holstein models. We will assume that we
are only working with one parabolic-band electron (or hole) so that the self-interaction functional depends on the electron position only through the term
$e^{i \vb{q} \cdot (\vb{r}(\tau) - \vb{r}(\tau'))}$. Provided this is true, we
can use Feynman's variational path integral method, and a form of the FHIP response
theory.

\section{Extending the Feynman variational polaron theory}

\subsection{Variational theory for a general polaron self-interaction functional}

Following the procedure for the Fr\"ohlich model~\cite{Feynman1955,Martin2023},
we derive a variational inequality for a general polaron, clearly moving the
model-specific evaluation into the self-interaction function. 
Our main constraint is that we cannot confine the electron to a discrete
lattice site in this theory. 
Even with this enforced continuum approximation, we find later that typical
features of the small lattice polaron are still present in the theory. 

The variational method for the polaron developed by Feynman gives a lower
upper-bound to the polaron free energy, 
\begin{equation}\label{eqn:general-feynman-jensen}
    \begin{aligned}
         F &\leq F_0(\beta) - \frac{1}{\hbar\beta} \langle S_{\text{pol}} - S_{0} \rangle_0 , \\
         &\leq F_0(\beta) -\frac{1}{\hbar^2\beta} \int_0^{\hbar\beta} d\tau \int_0^{\hbar\beta} d\tau'\ D_{\omega_0}(\tau - \tau') \\
         &\qquad\qquad\qquad\qquad\qquad\qquad\times \langle \Phi_{\text{pol}} - \Phi_0 \rangle_0 ,
    \end{aligned}
\end{equation}
where the expectation $\langle \mathcal{O} \rangle_0$ is defined as 
\begin{equation}
    \langle \mathcal{O} \rangle_0 \equiv \frac{\int \mathcal{D}^3 r(\tau)
    \mathcal{O} e^{-S_0[\vb{r}(\tau)]}}{\int \mathcal{D}^3 r(\tau)
    e^{-S_0[\vb{r}(\tau)]}}. 
\end{equation}
Here $S_0[\vb{r}(\tau)]$ is a trial action that is chosen to best approximate
the polaron model-action $S_{\text{pol}}[\vb{r}(\tau)]$, with the requirement
that the path integral for $S_0$ can be analytically evaluated.  
The trial-action is typically chosen to be at most quadratic in the electron
coordinate $\vb{r}(\tau)$ for this reason. 
We use the original 1955 quasi-particle trial action of Feynman, 
\begin{equation}
    \begin{aligned}
        &S_{0}[\vb{r}(\tau)] = \frac{m_b}{2} \int_0^{\hbar\beta} d\tau\ \Dot{\vb{r}}^2(\tau) \\
        &+ \frac{1}{\hbar} \int_0^{\hbar\beta} d\tau \int_0^{\hbar\beta} d\tau'\ D_{w}(\tau - \tau') \Phi_0\left[\vb{r}(\tau), \vb{r}(\tau')\right] ,
    \end{aligned}
\end{equation}
where the trial self-interaction functional is a simple quadratic,
\begin{equation}
    \Phi_0[\vb{r}(\tau), \vb{r}(\tau')] = \hbar w \kappa \left[\vb{r}(\tau) - \vb{r}(\tau')\right]^2 .
\end{equation}

The $\kappa$ and $w$ variational parameters have a direct interpretation as the
spring-constant and oscillation frequency of a quasi-particle. 
We have integrated out the interaction with the phonon-field, and replaced it
with a fictitious mass coupled to our electron by a spring, representing the
phonon drag. 
This trial model (the quasi-particle) is often reparameterised in terms of $v$
and $w$ variational parameters where $\kappa = m_b (v^2 - w^2)$. 

The expectation value of the trail action $\langle S_0 \rangle_0$ and the free energy of the trial system $F_0(\beta) $ are as given by \=Osaka~\cite{Osaka1959},
\begin{equation}
    \langle S_0 \rangle_0 = \frac{n\hbar\beta}{4} \frac{v^2-w^2}{v} \left(\frac{2}{v\hbar\beta} - \coth\left(\frac{v\hbar\beta}{2}\right)\right) ,
\end{equation}
\begin{equation}
    F_0(\beta) = \frac{n}{\beta} \log\left(\frac{w \sinh(\hbar\beta v / 2)}{v \sinh(\hbar\beta w / 2)}\right).
\end{equation}

All the expectation values in the variational expression can be evaluated from
$\langle e^{i \vb{q} \cdot (\vb{r}(\tau) - \vb{r}(\tau')} \rangle_0$, which for the trial model has a closed-form expression 
\begin{equation}
    \langle e^{i \vb{q} \cdot (\vb{r}(\tau) - \vb{r}(\tau'))} \rangle_0 = \exp\left[-\hbar q^2 G(\tau - \tau') / 2 m_b \right] ,
\end{equation}
where the imaginary-time thermal polaron Green's function $G(\tau)$ is given by
\begin{equation} \label{eqn:polarongreensfunc}
    \begin{aligned}
        G(\tau) &= \tau \left(1 - \frac{\tau}{\hbar\beta}\right) \\
        &+ \frac{v^2 - w^2}{v^3} \left[ D_v(0) - D_v(\tau) - v \tau \left(1 - \frac{\tau}{\hbar\beta} \right) \right],
    \end{aligned}
\end{equation}
where $D$ is the phonon propagator from Eqn.~(\ref{eqn:phonongf}).

In $n$-dimensions, from Eqn.~(\ref{eqn:general-feynman-jensen}) we have 
\begin{equation}\label{eqn:generalfreeenergy}
    \begin{aligned}
        F &\leq F_0(\beta) + \frac{1}{\beta} \langle S_0 \rangle_{0} \\
        &- \frac{2}{\hbar} \sum_{\vb{q}} \abs{V_{\vb{q}}}^2 \int_0^{\hbar\beta} d\tau\ D_{\omega_0}(\tau)\ e^{-\hbar q^2 G(\tau) / 2 m_b}
    \end{aligned}
\end{equation}
where we have used,
\begin{equation}
    \int_0^{\hbar\beta} d\tau \int_0^{\hbar\beta} d\tau'\ f(\abs{\tau - \tau'}) \sim 2\hbar\beta \int_0^{\hbar\beta}d\tau\ f(\tau),
\end{equation}
which is valid when the Hamiltonian for the system is time-translation invariant and $\beta$ is large.

\subsection{Fr\"ohlich Polaron Energy}

For the Fr\"ohlich self-interaction functional we have
\begin{equation}
    \begin{aligned}
        \langle\Phi_F\rangle_0 &= \frac{g^2_F \abs{S^{n-1}}}{(2\pi)^n} \int_0^{\infty} dq\ e^{-\hbar q^2 G(\tau) / 2 m_b} , \\
        &= \alpha_F \hbar^2 \omega_0^{3/2} \frac{\Gamma(\frac{n-1}{2})}{2 \Gamma(\frac{n}{2})} \frac{1}{\sqrt{G(\tau)}}.
    \end{aligned}
\end{equation}
where we have transformed the $q$-space summation into a spherical integral
over the $n$-dimensional ball 
\begin{equation} \label{eqn:general_self_interaction}
    \begin{aligned}
        \langle \Phi_{\text{pol}}\rangle_0 &= \sum_{\vb{q}} \abs{V_{\vb{q}}}^2 e^{-\hbar q^2 G(\tau) / 2 m_b} , \\
        &= \frac{V \abs{S^{n-1}}}{(2\pi)^n} \int_0^R dq \abs{V_q}^2 q^{n-1} e^{-\hbar q^2 G(\tau) / 2 m_b} ,
    \end{aligned}
\end{equation}
with $\abs{S^{n-1}} = 2\pi^{n/2}/\Gamma(n/2)$ the hypervolume of the unit $(n-1)$-sphere and $R$ the radius of the ball. 

The variational inequality for the Fr\"ohlich model is 
\begin{equation}
    \begin{aligned}
        F_F &\leq F_0(\beta) -\frac{1}{\beta} \langle S_0 \rangle_0 \\
        &- \alpha_F \hbar \omega_0^{3/2} \frac{\Gamma\left(\frac{n-1}{2}\right)}{\Gamma(\frac{n}{2})} \int_0^{\hbar\beta} d\tau\ \frac{D_{\omega_0}(\tau)}{\sqrt{G(\tau)}} .
    \end{aligned}
\end{equation}

\subsection{Holstein Polaron Energy}

For the parabolic Holstein model with a hypercubic lattice (i.e. cubic in 3D),
the self-interaction functional is  
\begin{equation}
    \begin{aligned}
        \langle \Phi_H \rangle_0 &= g^2_H \left[\frac{a}{2\pi} \int_{-\frac{\pi}{a}}^{\frac{\pi}{a}} dq\ e^{-\hbar q^2 G(\tau) / 2 m_b}\right]^n , \\
        &= 2 n \alpha_H J \hbar \omega_0 \left[\frac{\text{erf}\left(\pi \sqrt{G(\tau) J/\hbar} \right)}{\sqrt{4\pi G(\tau)J/\hbar}}\right]^n
    \end{aligned}
\end{equation}
where $\gamma = \hbar\omega_0 / J$ is the adiabaticity. 

Substituting the Holstein self-interaction functional into Eqn.~(\ref{eqn:general-feynman-jensen}) gives the variational inequality for the Holstein model as 
\begin{equation}
    \begin{aligned}
        F_H &\leq F_0(\beta) + \frac{1}{\beta} \langle S_0 \rangle_0  \\
        &- n \alpha_H J \omega_0 \int_0^{\hbar\beta} d\tau\ D_{\omega_0}(\tau) \Biggl[\frac{\text{erf}(\pi \sqrt{G(\tau)J/\hbar})}{\sqrt{4\pi G(\tau)J/\hbar}}\Biggr]^n
    \end{aligned}
\end{equation}
A key difference between the Holstein and Fr\"ohlich models is the domain of
the reciprocal-space integral.  For the Fr\"ohlich model this is over all of
reciprocal space and has spherical symmetry, whereas the Holstein model
integral is bounded (by the error function) to the first Brillouin Zone (and
formally reflects the crystal symmetry). Physically, this is an ultraviolet momentum
cutoff due to the discrete lattice in the Holstein model, without which the
integrals (in all spatial dimensions) catastrophically diverge. 
Additionally, the Fr\"ohlich model diverges in 1D whereas the Holstein model converges for all dimensionalities.

\section{Extending the FHIP Polaron Mobility Theory}

The polaron DC mobility may be obtained as for the Fr\"ohlich model in FHIP~\cite{Feynman1962}, from the real component of the frequency- and temperature-dependent conductivity function,
\begin{equation}\label{eqn:mobility}
    \mu_{dc} = \lim_{\Omega \to 0} \Re{\sigma(\Omega)} = \lim_{\Omega \to 0}
    \Re{\frac{1}{z(\Omega)}}. 
\end{equation}
Here the impedance function $z(\Omega)$ can be expressed in terms of the memory function $\Sigma(\Omega)$,
\begin{equation}
    z(\Omega) = i \left( \Omega - \Sigma(\Omega) \right).
\end{equation}
The memory function encodes the non-local-time interaction between the electron at different times. 

We can directly express the inverse DC mobility as the imaginary component of the memory function,
\begin{equation}
    \mu_{dc}^{-1} = \lim_{\Omega \to 0} \Im{\Sigma(\Omega)}.
\end{equation}
We first seek an expression for the dynamical memory function for a general
polaron, and delay specialising to the Holstein and Fr\"ohlich models until
later. 

\subsection{General Polaron Mobility}

The general memory function can be written as~\cite{Peeters1986}
\begin{equation}
    \begin{aligned}
        \Sigma(\Omega) &= \frac{2}{n m_b \hbar\Omega} \int_0^{\infty} dt\ \left(1 - e^{i \Omega t}\right) \\
        &\times\Im \left\{ \sum_{\vb{q}} \abs{V_{\vb{q}}}^2 q^2 D_{\omega_{\vb{q}}}(t) \langle e^{i \vb{q} \cdot \left[ \vb{r}(t) - \vb{r}(0) \right]} \rangle \right\}.
    \end{aligned}
\end{equation}
where we assume the rotational invariance of $\abs{V_{\vb{q}}}^2$ and $\omega_{\vb{q}}$. Here $D_{\omega_{\vb{q}}}(t)$ is the \emph{real}-time thermal and
\emph{dynamical} phonon Green's function,
\begin{equation}\label{eqn:real-time-phonon-gf}
    D_{\omega_{\vb{q}}}(t) = \coth(\frac{\hbar \beta \omega_{\vb{q}}}{2}) \cos(\omega_{\vb{q}} t) - i \sin(\omega_{\vb{q}} t),
\end{equation}
which can be obtained by substituting $\tau \to i t$ into Eqn.~(\ref{eqn:phonongf}). 

The key term to evaluate in the memory function is the density-density correlation function or dynamical structure factor 
\begin{equation}
    S_{\vb{q}}(t) = \langle \rho_{\vb{q}}(t) \rho^*_{\vb{q}}(0) \rangle = \langle e^{i \vb{q} \cdot \left[ \vb{r}(t) - \vb{r}(0) \right]} \rangle.
\end{equation}
This expectation value can be expressed as a path integral of the model action (Eqn.~(\ref{eqn:eph-action})) with an additional force term $f(t)=i \vb{q} \cdot \left[ \vb{r}(t) - \vb{r}(0) \right]$. However we would not be able to evaluate this exactly. 

Instead, we follow FHIP~\cite{Feynman1962} and insert the simple dynamic structure factor of the
Feynman (spring-mass) trial model, 
\begin{equation}
    \langle \rho_{\vb{q}}(t) \rho^*_{\vb{q}}(0) \rangle_0 = e^{-\hbar q^2 G(t) / 2 m_b},
\end{equation}
where $G(t)$ is the polaron Green function evaluated in real-time (i.e. substitute $\tau \to it$ into Eqn.~(\ref{eqn:polarongreensfunc})),
\begin{equation}
    \begin{aligned}
        G(t) &= i t \left(1 - \frac{i t}{\hbar \beta} \right) \\
        &+ \frac{v^2 - w^2}{v^3} \left[ D_v(0) - D_v(t)  - i v t \left(1 - \frac{i t}{\hbar \beta} \right) \right].
    \end{aligned}
\end{equation}
where $D$ is the phonon propagator from Eqn.~(\ref{eqn:real-time-phonon-gf}).

For a general polaron model, we thereby have the simplified memory function 
\begin{equation}
    \begin{aligned}
        \Sigma(\Omega) &= \frac{2}{n m_b \hbar\Omega} \int_0^{\infty} dt\ \left(1 - e^{i \Omega t}\right) \\
        &\times\Im \left\{ \sum_{\vb{q}} \abs{V_{\vb{q}}}^2 q^2 D_{\omega_{\vb{q}}}(t) e^{-\hbar q^2 G(t) / 2 m_b} \right\}.
    \end{aligned}
\end{equation}
To obtain the general polaron DC mobility, we take the zero frequency limit,
\begin{equation}
    \lim_{\Omega \to 0} \frac{\left(1 - e^{i \Omega t}\right)}{\Omega} \to -i t,
\end{equation}

In summary the general polaron DC mobility is 
\begin{equation}
    \begin{aligned}
         \mu_{dc}^{-1} = \frac{2}{n e \hbar} \int_0^\infty dt\ t \Im \left\{ \sum_{\vb{q}} \abs{V_{\vb{q}}}^2 q^2 D_{\omega_{\vb{q}}}(t) e^{-\hbar q^2 G(t) / 2 m_b} \right\}.
    \end{aligned}
\end{equation}

\subsection{Fr\"ohlich Polaron Mobility}

Equipped with the general polaron variational equations for the free energy and the corresponding memory function, we can specialise to the Fr\"ohlich model by evaluating the $q$-space integral, 
\begin{equation}
    I_F \equiv g_F^2 D_{\omega_0}(t) \int \frac{d^nq}{(2\pi)^n} \frac{q^2}{q^{n-1}} e^{-\hbar q^2 G(t) / 2 m_b}
\end{equation}
which we evaluate using spherical coordinates,
\begin{equation}
    \begin{aligned}
    I_F &= g^2_F D_{\omega_0}(t) \frac{\abs{S^{n-1}}}{(2\pi)^n} \int_0^\infty dq\ q^2 e^{-\hbar q^2 G(t) / 2 m_b}, \\
    &= g_F^2 D_{\omega_0}(t) \frac{\abs{S^{n-1}}}{(2\pi)^n} \frac{2\pi^2}{(2\pi \hbar G(t) / m_b)^{3/2}}, \\
    &= \alpha_F m_b \hbar\omega_0^{3/2} \frac{\Gamma(\frac{n-1}{2})}{\Gamma(\frac{n}{2})} \frac{D_{\omega_0}(t)}{G(t)^{3/2}} .
    \end{aligned}
\end{equation}
This gives the memory function for the Fr\"ohlich model as 
\begin{equation}
    \Sigma_F(\Omega) = \frac{2 \alpha_F \omega_0^{3/2}}{n} \frac{\Gamma(\frac{n-1}{2})}{\Gamma(\frac{n}{2})} \int_0^\infty dt\ \frac{\left(1 - e^{i \Omega t}\right)}{\Omega} \Im \left\{ \frac{D_{\omega_0}(t)}{G(t)^{3/2}} \right\}.
\end{equation}
The inverse Fr\"ohlich DC mobility is 
\begin{equation}
    \mu_{F}^{-1} = \frac{2 \alpha_F m_b \omega_0^{3/2}}{n e} \frac{\Gamma(\frac{n-1}{2})}{\Gamma(\frac{n}{2})} \int_0^\infty dt\ \frac{t D_{\omega_0}(t)}{ G(t)^{3/2}}.
\end{equation}
These expressions for mobility are equivalent to those proposed in Eqns. (35) and (41) in FHIP~\cite{Feynman1962}, generalised to arbitrary dimension and with the zero-frequency taken explicitly.

\subsection{Holstein Polaron Mobility}

We specialise to the Holstein model by evaluating the $q$-space integral,
\begin{equation}
    I_H \equiv g_H^2 D_{\omega_0}(t) a^n \int \frac{d^nq}{(2\pi)^n} q^2 e^{-\hbar q^2 G(t) / 2 m_b}
\end{equation}
which, assuming the crystal to be hypercubic, we evaluate as,
\begin{equation}
    \begin{aligned}
    I_H &= n g^2_H D_{\omega_0}(t) (a/2\pi)^n \int_{-\frac{\pi}{a}}^{\frac{\pi}{a}} dq_{\paral}\ q_{\paral}^2\ e^{-\hbar q_{\paral}^2 G(t) / 2 m_b} \\
    &\times \left[ \int_{-\frac{\pi}{a}}^{\frac{\pi}{a}} dq_{\perp}\ e^{-\hbar q_{\perp}^2 G(t) / 2 m_b} \right]^{n-1}, \\
    &= g^2_H D_{\omega_0}(t) \frac{n}{a^2} \left[ \frac{\text{erf}(\pi \sqrt{G(t)J/\hbar})}{2\sqrt{\pi G(t)J/\hbar}} \right]^{n-1} \\
    &\times \left[ \frac{\text{erf} (\pi \sqrt{G(t)J/\hbar})}{4 \sqrt{\pi} (G(t)J/\hbar)^{3/2}} - \frac{e^{-\pi^2 G(t)J/\hbar}}{2 G(t)J/\hbar} \right] \\ 
    &= \frac{n^2 \alpha_H m_b J^2 \omega_0}{2\hbar} D_{\omega_0}(t) \left[ \frac{\text{erf}(\pi \sqrt{G(t)J/\hbar})}{2\sqrt{\pi G(t)J/\hbar}} \right]^{n-1} \\
    &\times \left[ \frac{\text{erf} (\pi \sqrt{G(t)J/\hbar})}{\sqrt{\pi} (G(t)J/\hbar)^{3/2}} - \frac{2 e^{-\pi^2 G(t) J/\hbar}}{G(t) J/\hbar} \right] .
    \end{aligned}
\end{equation}
The memory function for the Holstein model is, 
\begin{equation}
    \begin{aligned}
        \Sigma_H(\Omega) &= \frac{2 n^2 \alpha_H J^2 \omega_0}{\hbar^2} \int_0^\infty dt\ \frac{\left(1 - e^{i \Omega t}\right)}{\Omega} \Im \Biggl\{ D_{\omega_0}(t) \\
        &\times \left[ \frac{\text{erf}(\pi \sqrt{G(t)J/\hbar})}{2\sqrt{\pi G(t)J/\hbar}} \right]^{n-1} \\
        &\times \left[ \frac{\text{erf} (\pi \sqrt{G(t)J/\hbar})}{\sqrt{\pi} (G(t)J/\hbar)^{3/2}} - \frac{2 e^{-\pi^2 G(t) J/\hbar}}{G(t) J/\hbar} \right] \Biggr\}.
    \end{aligned}
\end{equation}
The inverse Holstein DC mobility is 
\begin{equation}
    \begin{aligned}
        \mu_{H}^{-1} &= \frac{2 n^2 \alpha_H m_b J^2 \omega_0}{e \hbar^2} \int_0^\infty dt\ t D_{\omega_0}(t) \\
        &\times \left[ \frac{\text{erf}(\pi \sqrt{G(t)J/\hbar})}{2\sqrt{\pi G(t)J/\hbar}} \right]^{n-1} \\
        &\times \left[ \frac{\text{erf} (\pi \sqrt{G(t)J/\hbar})}{\sqrt{\pi} (G(t)J/\hbar)^{3/2}} - \frac{2 e^{-\pi^2 G(t) J/\hbar}}{G(t)J/\hbar} \right].
    \end{aligned}
\end{equation}

A notable issue for the numerics is that we have error functions within our
integrals.

\section{Results}

We implement the previously described methods in our open-source codes
\textsc{PolaronMobility.jl}~\cite{JOSS, GITHUB}. 
Using these codes we first provide results for the abstract Holstein model, and compare these results to Ragni's Diagrammatic Monte Carlo results, and the Fr\"ohlich Hamiltonian. 
We look at one- to three-dimensional models. 
The Fr\"ohlich model, is
of mainly academic interest as most continuum materials are fairly isotropic. 
For the Holstein model, organic semiconductors are often highly anisotropic,
and the varying behaviour is of direct technical interest. 
We then predict the polaronic behaviour of the organic crystalline semiconductor Rubrene. 

To characterise the the models, we will consider:
\begin{enumerate}
    \item The energy and character of the athermal quasi-particle (polaron) state versus coupling.
    \item The temperature dependence of the polaron mobility, and polaron energy.
    \item The frequency dependence of the polaron mobility, and optical conductivity. 
\end{enumerate}

We note that the literature for the Holstein model is inconsistent in the
conventions of defining the parameters of the Holstein model, especially how
these parameters depend on the number of spatial dimensions (or not). 
In that sense, it is probably more universal and clear to refer to the models
\emph{unitless} parameters.
For the Fr\"ohlich model this is the electron-phonon coupling strength
$\alpha$. 
For the Holstein model, as well as the equivalent $\alpha_H$, the second energy
scale (of hopping integral versus phonon energy) requires an additional 
adiabaticity parameter $\gamma$. 
We follow the convention used by Ragni~\cite{Ragni2020} in their Diagrammatic
Monte Carlo (DiagMC) study, which we plot our results against. 

%Aside from the new variational approximation to the Holstein model, in this paper, we will also present (alongside the Holstein data) three-dimensional Fr\"ohlich polaron data so that the two models' differences are apparent.

Lastly, we state what is absent: we do not have any numeric results for the
general k-space integration form of the theory. 
This involves much more involved computation, and ingesting a suitably
validated set of electronic structure electron-phonon calculations. 
Work in this direction is ongoing. 

\subsection{Athermal polarons}

We start with the zero temperature polaron in one-, two- and three dimensions for the parabolic Holstein model and the three-dimensional Fr\"ohlich model. We present the models in a unitless presentation where the adiabaticity $\gamma = 1$ ($\omega_0 = J = 1$). We also assume that two dimensionless coupling alpha parameters, $\alpha_H$ and $\alpha_F$, used in the parabolic Holstein and Fr\"ohlich models produce similar physical regimes within either model for a point of comparison. Therefore we will refer to a single alpha parameter $\alpha \equiv \alpha_H = \alpha_F$ when comparing these two models.

\subsubsection{Polaron Variational Parameters}

\begin{figure}[!t]
    \includegraphics[width=0.49\textwidth]{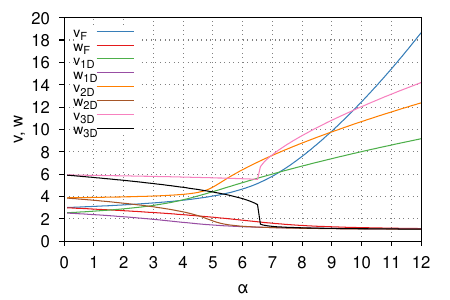}
    \caption{Optimal values of the polaron variational parameters $v$ and $w$ for the three-dimensional Fr\"ohlich and one-, two- and three-dimensional Holstein models with respect to the electron-phonon dimensionless coupling parameter $\alpha$. Here we have $J = \omega_0 = 1$.}
    \label{fig:vw_alpha}
\end{figure}

In Fig.~(\ref{fig:vw_alpha}) are the polaron variational parameters $v$ and $w$ of the Holstein and Fr\"ohlich polarons as a function of the electron-phonon dimensionless coupling parameter $\alpha$. The Holstein model has a noticeably different variational solution than the Fr\"ohlich model, with a distinct discontinuity in three dimensions. This transition is smoother for one- and two-dimensions and occurs at $\alpha
\approx 2n$, where $n$ is the dimensionality of the model. We interpret this transition as physically corresponding to the formation of a small-polaron state.
 
In both models, $w$ asymptotes to  $\omega_0 = 1$ at large electron-phonon coupling $\alpha$, with a more abrupt transition in the Holstein model. The $v$ parameters have a different strong coupling dependency on $\alpha$. In the Fr\"ohlich model at large $\alpha$, $v_F \sim \alpha^2$ whereas in the Holstein model $v_H \sim \sqrt{\alpha}$. Another noticeable difference is the weak coupling limit ($\alpha \to 0$) where the Fr\"ohlich model parameters asymptotes are $v_F = w_F = 3$, whereas Holstein are $v_H = w_H \approx 2n$.

\subsubsection{Polaron Ground-state Energy}

\begin{figure}[!t]
    \includegraphics[width=0.49\textwidth]{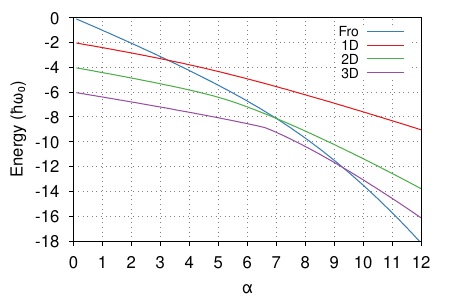}
    \caption{The polaron ground-state (athermal) energy of the three-dimensional Fr\"ohlich and one-, two- and three-dimensional Holstein models with respect to the electron-phonon coupling parameter $\alpha$. Here we have $J = \omega_0 = 1$.}
  \label{fig:E_alpha}
\end{figure}

\begin{figure*}
  \begin{subfigure}[b]{0.49\textwidth}
    \includegraphics[width=\textwidth]{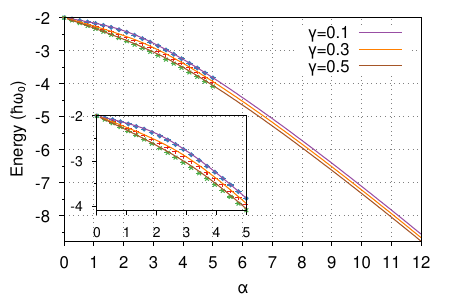}
  \end{subfigure}
  \hfill
  \begin{subfigure}[b]{0.49\textwidth}
    \includegraphics[width=\textwidth]{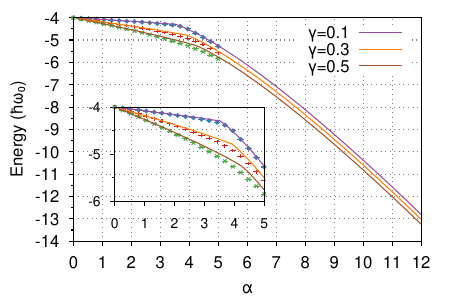}
  \end{subfigure}
  \begin{subfigure}[b]{0.5\textwidth}
    \includegraphics[width=\textwidth]{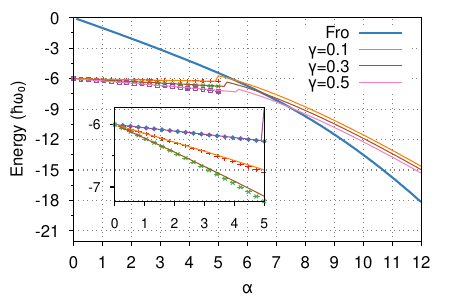}
  \end{subfigure}
  \hfill
  \begin{subfigure}[b]{0.49\textwidth}
    \includegraphics[width=\textwidth]{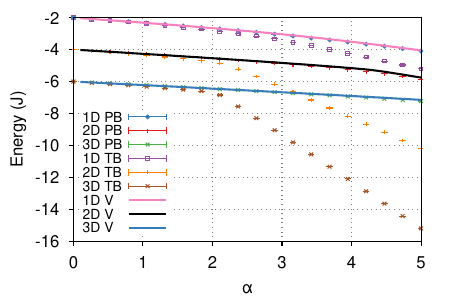}
  \end{subfigure}
  \caption{Polaron binding energy for the parabolic Holstein model with respect to the electron-phonon dimensionless coupling parameter $\alpha_H$ for dimensions $n = 1$, $2$ and $3$, and adiabaticities $\gamma = 0.1$, $0.3$ and $0.5$. Here we show both the results of our variational model and DiagMC data provided by Ragni for $\alpha = 0$ to $5$ (shown more closely in the insets). \textbf{Top Left:} One-dimensional comparison with DiagMC data. \textbf{Top Right:} Two-dimensional comparison with DiagMC data. \textbf{Bottom Left:} Three-dimensional comparison with DiagMC data. We also include the variational solution for the three-dimensional Fr\"ohlich model (with $\omega_0 = 1$). \textbf{Bottom Right:} Comparison of parabolic Holstein model with the full tight-binding Holstein model DiagMC results for $\gamma = 0.5$. Here we also show the variational solution for the parabolic model.}
  \label{fig:energy_alpha}
\end{figure*}

In Fig.~(\ref{fig:E_alpha}) is the variational solution for the free energy of the one-, two- and three-dimensional parabolic Holstein model (with $J = \omega_0 = 1$ in this unitless presentation) with respect to the electron-phonon coupling parameter $\alpha$. For comparison, we also present the variational solution to the three-dimensional Fr\"ohlich model ($\omega_0 = 1$).

The athermal polaron energy at $\alpha = 0$ corresponds to the band extrema for either model. In the Fr\"ohlich model this is zero $E_F(\alpha=0) = 0$ and for the Holstein model, this is $E_H(\alpha=0) = 2nJ$ with $n$ the dimensionality. The athermal energy in the Fr\"ohlich model is approximately linear for small alpha $E_F \sim -\hbar\omega_0\alpha$ and quadratic for large alpha $E_F \sim -\hbar\omega_0 \alpha^2$. The athermal energy in the parabolic Holstein model is also linear for small alpha $E_H + 2nJ \sim  -J\alpha/2$ (which is the same for all dimensions), but the large alpha behaviour is not quadratic and is instead linear and dependent on the number of dimensions $E_H +2nJ \sim -J n \alpha$.

In Fig.~(\ref{fig:energy_alpha}) we now vary the adiabaticity $\gamma = \hbar\omega_0/J = 0.1, 0.3$ and $0.5$ in the parabolic Holstein model for dimensions $n = 1, 2$ and $3$ and compare our variational results with diagrammatic Monte-Carlo (diagMC) results provided by Ragni (for values of $\alpha = 0$ to $5$ which are shown more closely in the inset figures). Our variational solutions show good agreement with Ragni's diagMC results for all dimensions and adiabaticities. The adiabaticity affects the lower alpha energy of the polaron below the small polaron transition around $\alpha = 2n$ and increases the sharpness of this transition for smaller adiabaticity. We note that in three dimensions with $\gamma < 1$ this transition momentarily \emph{reduces} the polaron binding energy in the variational solution, which is not replicated in the diagMC results --  more diagMC data for $\alpha > 5$ may be required to determine if this is an artefact of the parabolic Holstein model or just the variational solution. In Fig.~(\ref{fig:energy_alpha}) we also show (bottom-right sub-figure) a comparison of the diagMC results for the \emph{parabolic-band} Holstein model with the original \emph{tight-binding-band} Holstein model where we see that the two models predict similar energies below the small-polaron transition, but the tight-binding model shows a sharper, dimension-independent transition at $\alpha = 2$ and larger polaron-binding energy above this transition. 

\subsection{Thermal polarons}

We will now look at the finite temperature dependence of the polaron in two- and three-dimensions for the parabolic Holstein model (top sub-figures) and the three-dimensional Fr\"ohlich model (bottom sub-figure), for values of the electron-phonon coupling $\alpha = 0.1, 2, 4, 6, 8, 10, 12$. Here we also take a unitless presentation with $\omega_0 = J = 1$. 

\subsubsection{Polaron Free Energy}

\begin{figure*}[!tbp]
  \begin{subfigure}[b]{0.49\textwidth}
    \includegraphics[width=\textwidth]{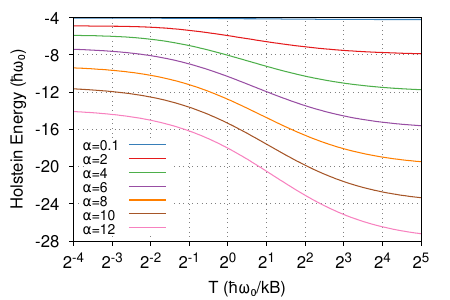}
  \end{subfigure}
  \hfill
  \begin{subfigure}[b]{0.49\textwidth}
    \includegraphics[width=\textwidth]{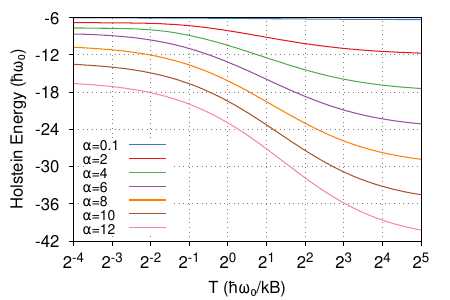}
  \end{subfigure}
  \begin{subfigure}[b]{0.49\textwidth}
    \includegraphics[width=\textwidth]{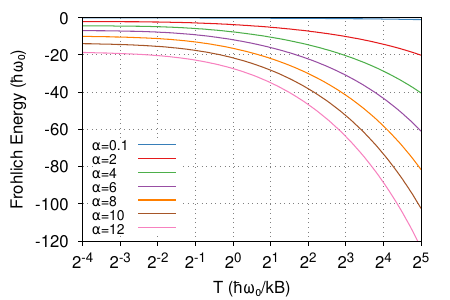}
  \end{subfigure}
  \caption{Temperature dependence ($T$, in units of phonon frequency $\hbar\omega_0/k_B$) of the polaron free energy for the parabolic Holstein model in two-dimensions (top-left) and three-dimensions (top-right) and for the Fr\"ohlich model in three-dimensions (bottom), for values of the electron-phonon coupling $\alpha = 0.1, 2, 4, 6, 8, 10, 12$. Here $\omega_0 = J = 1$.}
  \label{fig:energy_temp}
\end{figure*}

In Fig.~(\ref{fig:energy_temp}) is the temperature dependence of the polaron free energy.

In the parabolic Holstein model, the polaron-free energy transitions from the ground-state energy at very low temperatures to  $E_H = -2nJ - \alpha n J$ at large temperatures with the transition point occurring around the Debye temperature $T_D = \hbar\omega_0/k_B$. Comparatively, whilst the Fr\"ohlich model also shows a transition around the Debye temperature, above this temperature the polaron free energy increases as $E_F \sim -T^{1/2}$ without bound.

\subsubsection{Polaron Mobility}

\begin{figure*}[!tbp]
  \begin{subfigure}[b]{0.49\textwidth}
    \includegraphics[width=\textwidth]{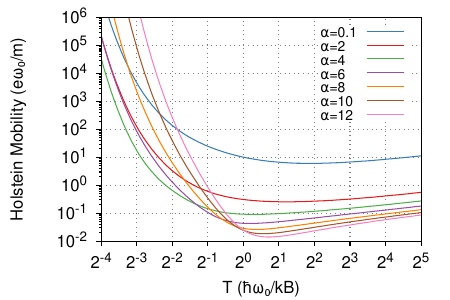}
  \end{subfigure}
  \hfill
  \begin{subfigure}[b]{0.49\textwidth}
    \includegraphics[width=\textwidth]{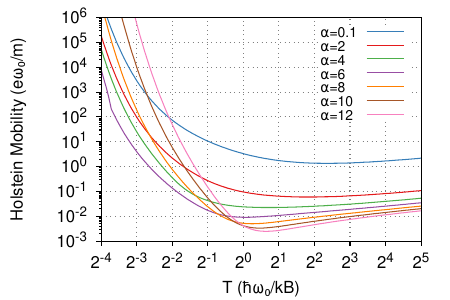}
  \end{subfigure}
  \begin{subfigure}[b]{0.49\textwidth}
    \includegraphics[width=\textwidth]{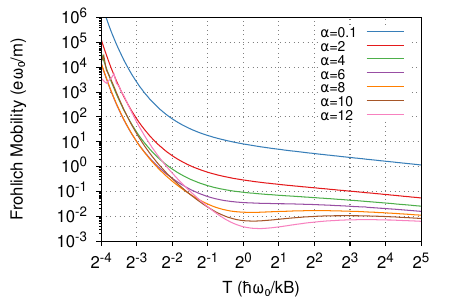}
  \end{subfigure}
  \caption{Temperature dependence ($T$, in units of phonon frequency $\hbar\omega_0/k_B$) of the polaron mobility for the parabolic Holstein model in two-dimensions (top-left) and three-dimensions (top-right) and for the Fr\"ohlich model in three-dimensions (bottom), for values of the electron-phonon coupling $\alpha = 0.1, 2, 4, 6, 8, 10, 12$. Here $\omega_0 = J = 1$.}
  \label{fig:mobility_temp}
\end{figure*}

In Fig.~(\ref{fig:mobility_temp}) we have the temperature dependence of the polaron mobility.

At weaker coupling, the mobility shows the typical exponentially decreasing band-like transport for temperatures below the Debye temperature $T_D = \hbar\omega_0/k_B$ for both the parabolic Holstein and Fr\"ohlich models. 

Above the Debye temperature, the temperature dependence transitions to a power-law relationship $T^{-x}$ where $x$ is some number typically used to determine the dominant scattering mechanism within a material. With a semi-classical theory, for acoustic phonons this is $x = 3/2$, for optical phonons, it is $x = 1/2$. For the Fr\"ohlich model, the high-temperature mobility follows the power-law $\mu_F \sim T^{-1/2}$ of optical phonons. However, the parabolic Holstein model follows a power-law of $\mu_H \sim T^{1/2}$ at high temperatures indicative of thermally-activated small polaron hopping. At low temperatures, the mobility of both models increases abruptly below the Debye temperature due to the increasing contribution of the (adiabatic) electron transfer without phonon participation.

As the electron-phonon coupling increases, we begin to see the onset of the ski-slope feature~\cite{Mishchenko2019, MartinMultiple2022} in the Fr\"ohlich model where the mobility takes on a local minimum at the Deybe temperature before increasing to a local maximum at the polaron quasiparticle frequency $T = \hbar v / k_B$ and then transitioning to the power-law relationship at higher temperatures. This ski-slope is absent in the parabolic Holstein model which has only the minimum at the Debye temperature which becomes deeper at larger alphas.

\subsection{Dynamic response}

In this section, we look at how the parabolic Holstein model varies with the frequency of an external, perturbing, varying electric field. Specifically, the polaron memory function which we compare to the Fr\"ohlich polaron results of FHIP~\cite{Feynman1962}, and the optical conductivity which we compare to the Fr\"ohlich polaron results of DSG~\cite{Devreese1972}. We look at frequencies $\Omega / \omega_0$ from $0$ to $30$ and $\alpha = 1, 2, 3, 4, 5, 6$ and $7$. 

\subsubsection{Polaron Memory Function}

\begin{figure*}
\centering
  \begin{subfigure}[b]{0.49\textwidth}
    \includegraphics[width=\textwidth]{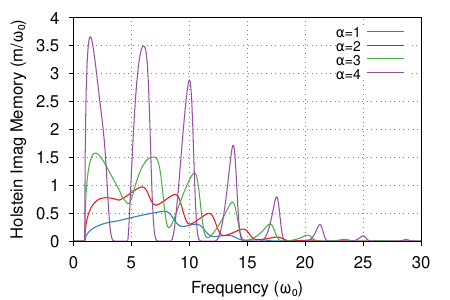}
  \end{subfigure}
  \hfill
  \begin{subfigure}[b]{0.49\textwidth}
    \includegraphics[width=\textwidth]{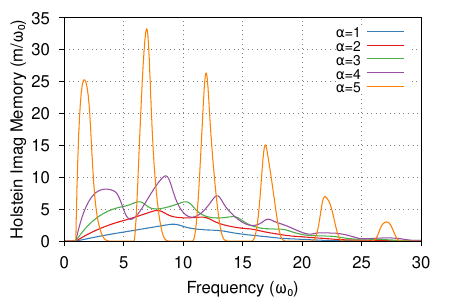}
  \end{subfigure}
  \begin{subfigure}[b]{0.49\textwidth}
    \centering
    \includegraphics[width=\textwidth]{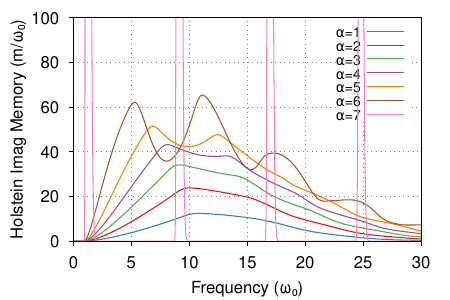}
  \end{subfigure}
  \hfill
  \begin{subfigure}[b]{0.49\textwidth}
    \centering
    \includegraphics[width=\textwidth]{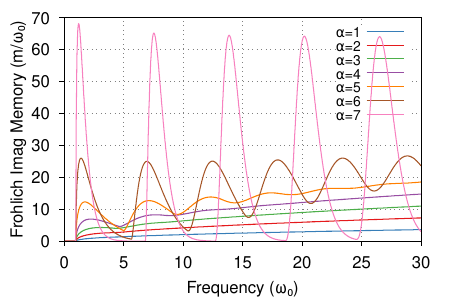}
  \end{subfigure}
  \caption{Frequency dependence ($\Omega$, in units of the phonon frequency $\omega_0$) of the polaron memory function $\chi(\Omega)$ for the parabolic Holstein model in one-dimension (top-left), two-dimensions (top-right) and three-dimensions (bottom-left) and for the Fr\"ohlich model in three-dimensions (bottom-right). We show the memory function for multiple values of the electron-phonon coupling $\alpha$ ranging from $1$ to $4$ in one-dimension, $5$ in two-dimension and $7$ in three-dimensions. These upper limits were chosen due to the small-polaron state being formed above $\alpha  =  2n$ where $n$ is the dimensionality. We only show the imaginary component of the memory function to reduce clutter. Here $m = \omega = J = 1$.}
  \label{fig:im_mem_freq}
\end{figure*}

Fig.~(\ref{fig:im_mem_freq}) is the frequency-dependent imaginary component of the memory function for the parabolic Holstein and Fr\"ohlich polarons for a range of electron-phonon coupling strengths. Note that here we use the FHIP~\cite{Feynman1962} definition of the memory function $\chi(\Omega)$, which is related to the memory function used by us and Devreese and Peeters~\cite{Peeters1986} $\Sigma(\Omega)$  by the expression $\Sigma(\Omega) = \chi(\Omega) / \Omega$. We choose the alpha values $1$ to $4$ for one dimension, $1$ to $5$ for two dimensions and $1$ to $7$ in three dimensions as above these limits the Holstein model enters the strongly-coupled small polaron regime and the corresponding memory functions limit to delta-function peaks at $\Omega_{\text{peaks}} = 1 + n v, n \in \mathbf{N}$ where $v_H$ is the Holstein polaron quasiparticle frequency. 

These peaks correspond to one- two- three- etc phonon excitations. At lower electron-phonon coupling strengths these peaks become more dispersed and damped until they eventually merge at small couplings. From Fig.~(\ref{fig:im_mem_freq}) we see that the Holstein memory function is larger for larger dimensionality and becomes more heavily weighted towards larger frequencies as the polaron quasiparticle frequency increases with dimension.

The Holstein memory function shows a Poisson-distributed envelope with a maximum at finite frequencies favouring one-phonon processes in one-dimension and two-phonon processes in two- and three-dimensions. Conversely, the three-dimensional Fr\"ohlich memory function assigns roughly equal weight to each phonon process and has an envelope function that slowly increases with frequency and has no apparent maximum. The bottom-right sub-figure in Fig.~(\ref{fig:im_mem_freq}) shows the Fr\"ohlich memory function and reproduces the results of $\alpha = 3,5$ and $7$ from figures 1-3 of FHIP~\cite{Feynman1962} as well as additionally the results for $\alpha = 1, 2, 4$ and $6$.

\subsubsection{Polaron Optical Conductivity}

\begin{figure*}[!tbp]
    \centering
  \begin{subfigure}[b]{0.49\textwidth}
    \centering
    \includegraphics[width=\textwidth]{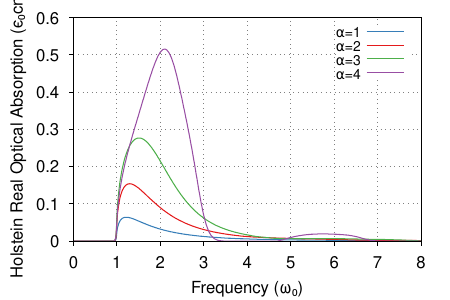}
  \end{subfigure}
  \hfill
  \begin{subfigure}[b]{0.49\textwidth}
    \centering
    \includegraphics[width=\textwidth]{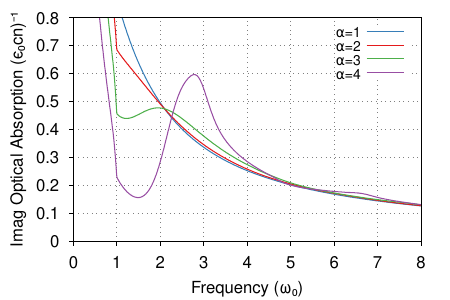}
  \end{subfigure}
  \begin{subfigure}[b]{0.49\textwidth}
    \centering
    \includegraphics[width=\textwidth]{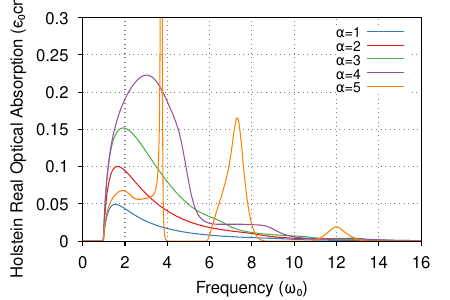}
  \end{subfigure}
  \hfill
  \begin{subfigure}[b]{0.49\textwidth}
    \centering
    \includegraphics[width=\textwidth]{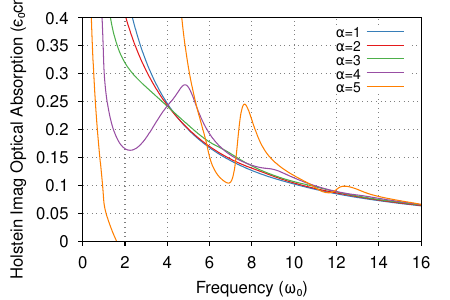}
  \end{subfigure}
  \begin{subfigure}[b]{0.49\textwidth}
    \centering
    \includegraphics[width=\textwidth]{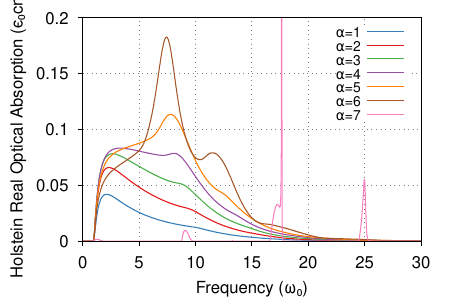}
  \end{subfigure}
  \hfill
  \begin{subfigure}[b]{0.49\textwidth}
    \centering
    \includegraphics[width=\textwidth]{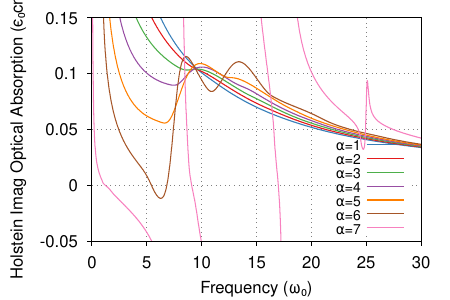}
  \end{subfigure}
  \begin{subfigure}[b]{0.49\textwidth}
    \centering
    \includegraphics[width=\textwidth]{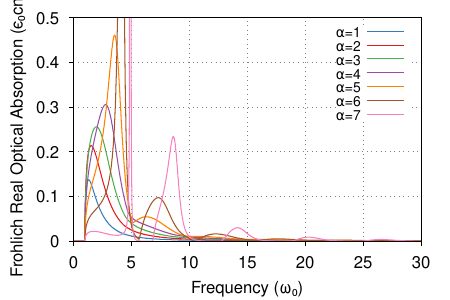}
  \end{subfigure}
  \hfill
  \begin{subfigure}[b]{0.49\textwidth}
    \centering
    \includegraphics[width=\textwidth]{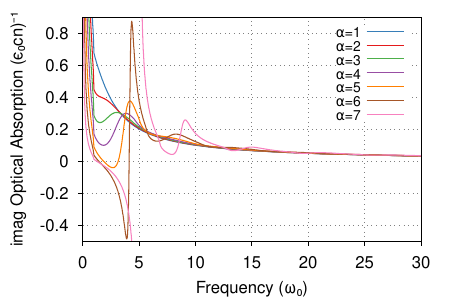}
  \end{subfigure}
  \caption{Frequency dependence ($\Omega$, in units of the phonon frequency $\omega_0$) of the complex conductivity $\sigma(\Omega)$ for the parabolic Holstein model in one-dimension (first row), two-dimensions (second row) and three-dimensions (third row) and the Fr\"ohlich model in three-dimensions (last row), for multiple electron-phonon couplings. We show the real component of the conductivity on the left and the imaginary component on the right. Here $m = \omega = J = 1$.}
  \label{fig:con_freq}
\end{figure*}

In Fig.~(\ref{fig:con_freq}) is the frequency-dependent complex conductivity (otherwise known as the optical conductivity) for the Holstein and Fr\"ohlich polarons for a range of electron-phonon coupling strengths. Electron-phonon coupling values $\alpha = 1, 3, 5, 6, 7$ are a direct comparison to figures (1-5) in DSG~\cite{Devreese1972} and the result of these figures are reproduced in the bottom-left sub-figure of Fig.~(\ref{fig:con_freq}). We show here both the real and imaginary components of conductivity for completeness but will focus our discussion on the real component. 

Starting at weak coupling $\alpha = 1, 2$, both models in all the presented spatial dimensions show the same form of a one-phonon excitation peak that decays away at higher frequencies. We see more structure at intermediate coupling $\alpha = 3, 4$. Firstly, the one-dimensional Holstein polaron shows a two-phonon peak that only just shows as a side-band in two dimensions and is barely visible merged with the one-phonon peak in three dimensions. For the three-dimensional Fr\"ohlich model only the one-phonon peak is visible at these couplings.

As we approach strong electron-phonon coupling $\alpha = 5$ we enter the small polaron regime for the one- and two-dimensional Holstein models. In two dimensions, the one-phonon peak develops a sharp delta function peak at $\Omega = 3.8 \omega_0$ and the two- and three-phonon peaks become more pronounced. This sharp peak is an intense Relaxed Excited State (RES) transition and the low-frequency shoulder represents the original one-phonon peak. The two- and three-phonon peaks are otherwise known as Frank-Condon states. In three dimensions, the Holstein model develops a RES on the high-frequency side of the one-phonon peak at $\Omega = 7 \omega_0$. At $\alpha = 5$, the Fr\"ohlich model begins to develop a RES merged with the one-phonon peak at $\Omega = 3.5 \omega_0$ and also the two-phonon peak starts to be visible at $\Omega - 6.2 \omega_0$.

At $\alpha = 6$, the three-dimensional Holstein model now has a more prominent RES peak and two-phonon peak at $\Omega = 12 \omega_0$. The three-phonon peak has also become visible. The Fr\"ohlich model develops similar features except that the multiple-phonon peaks are distinctly separated and the RES transition is sharper.

Finally, at $\alpha = 7$, we enter into the small-polaron regime for the three-dimensional Holstein model where the one- and two-phonon peaks are greatly diminished and the RES on the one-phonon peak has gone. The three- and four-phonon peaks are now more prominent and sharp, with a very intense RES on at $\Omega = 17.8 \omega_0$ on the three-phonon peak. Likewise, the Fr\"ohlich model sees a reduction of the one-phonon peak, but instead retains the intense RES and has the two-phonon peak be most prominent despite the three- and four-phonon peaks becoming barely visible.

\subsection{Real organic materials: Rubrene}

\begin{table*}
    \centering
    \begin{tabular}{|c|c|c|c|c|c|c|c|}
    \hline
        $g$ (meV) & $\omega_0$ (THz) & $J$ (meV) & $a$ (Å) & $\gamma$ & $m_b$ ($m_e$) & $\lambda^2$ & $\alpha$ \\
    \hline
         $106.8$ & $5.768$ & $134.0$ & $14.06$ & $0.178$ & $0.144$ & $19.75$ & $0.586$ \\
    \hline
    \end{tabular}
    \caption{3D Rubrene Bulk crystal data derived from~\cite{Ordejn2017}. Here $g$ is the Holstein hole-phonon coupling element, $\omega_0$ is the single-mode effective phonon frequency, $J$ is the electron transfer/hopping integral, $a$ is the geometric-meaned crystal lattice constant, $\gamma$ is the Holstein adiabaticity unitless parameter, $m_b$ is the effective hole band-mass, $\lambda^2 = (g / \hbar\omega_0)^2$ is the unitless squared hole-phonon coupling element and $\alpha = \lambda^2 \gamma / 6$ is the 3D unitless Holstein electron-phonon parameter.}
    \label{tab:rubrene_data}
\end{table*}

\begin{table*}
    \centering
    \begin{tabular}{|c|c|c|c|c|c|}
    \hline
        & $v_0$ (THz) & $w_0$ (THz) & $M_0$ ($m_e$) & $R_0$ (Å) & $F_0$ (meV) \\
    \hline
         Holstein & $21.01$ & $7.258$ & $0.146$ & $121.3$ & $-6.867$ \\
    \hline
         Fr\"ohlich & $17.66$ & $16.88$ & $0.149$ & $304.2$ & $-5.842$ \\
    \hline
    \end{tabular}
    \caption{Ground-state polaron properties for a Rubrene Bulk crystal calculated using the variational Holstein and Fr\"ohlich models.}
    \label{tab:gs_rubrene}
\end{table*}

\begin{table*}
    \centering
    \begin{tabular}{|c|c|c|c|c|c|c|}
    \hline
        & $v$ (THz) & $w$ (THz) & $M$ ($m_e$) & $R$ (Å) & $F$ (meV) & $\mu$ (cm$^2$V$^{-1}$s$^{-1}$) \\
    \hline
         Holstein & $77.22$ & $74.37$ & $0.155$ & $34.28$ & $-18.10$ & $47.72$ \\
    \hline
        Fr\"ohlich & $43.56$ & $41.15$ & $0.151$ & $99.10$ & $-10.95$ & $92.59$ \\
    \hline
    \end{tabular}
    \caption{Room temperature ($300$ K) polaron properties for a Rubrene Bulk crystal calculated using the variational Holstein and Fr\"ohlich models.}
    \label{tab:rt_rubrene}
\end{table*}

\begin{figure*}[!tbp]
    \centering
  \begin{subfigure}[b]{0.49\textwidth}
    \centering
    \includegraphics[width=\textwidth]{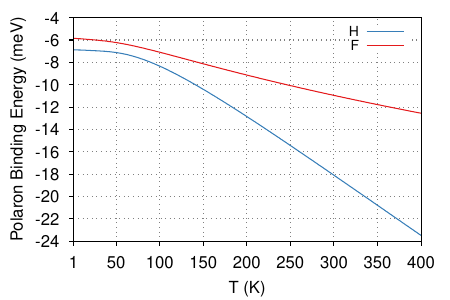}
    \label{fig:rubrene_F_temp}
  \end{subfigure}
  \hfill
  \begin{subfigure}[b]{0.49\textwidth}
    \centering
    \includegraphics[width=\textwidth]{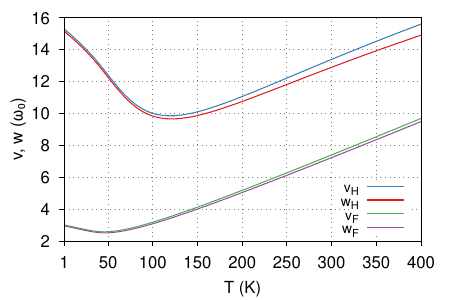}
    \label{fig:rubene_vw_temp}
  \end{subfigure}
  \begin{subfigure}[b]{0.49\textwidth}
    \centering
    \includegraphics[width=\textwidth]{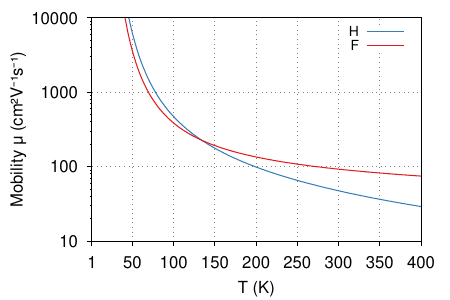}
    \label{fig:rubrene_temp}
  \end{subfigure}
  \caption{Polaron properties predicted from the variational parabolic Holstein and Fr\"ohlich models for a bulk 3D Rubrene organic crystal. \textbf{Top-left:} The polaron binding energy (the polaron self-energy relative to the band extrema) $F$ (\si{meV}) in Rubrene as a function of temperature (\si{K}). \textbf{Top-right:} Optimal variational parameters $v$ and $w$ (\si{THz}) as a function of temperature (K). \textbf{Bottom} The DC polaron mobility $\mu(T)$ (\si{cm^2V^{-1}s^{-1}}) as a function of temperature $T$ (\si{K}).}
  \label{fig:rubrene_temp}
\end{figure*}

% \begin{figure*}[!tbp]
%     \centering
%   \begin{subfigure}[b]{0.49\textwidth}
%     \centering
%     \includegraphics[width=\textwidth]{plots/rubrene/rubrene-frohlich-real-memory-temp-0.4to400K-freq-0to10omega-COLOUR.pdf}
%   \end{subfigure}
%   \hfill
%   \begin{subfigure}[b]{0.49\textwidth}
%     \centering
%     \includegraphics[width=\textwidth]{plots/rubrene/rubrene-frohlich-imag-memory-temp-0.4to400K-freq-0to10omega-COLOUR.pdf}
%   \end{subfigure}
%   \begin{subfigure}[b]{0.49\textwidth}
%     \centering
%     \includegraphics[width=\textwidth]{plots/rubrene/rubrene-holstein-real-memory-temp-0.4to400K-freq-0to10omega-COLOUR.pdf}
%   \end{subfigure}
%   \hfill
%   \begin{subfigure}[b]{0.49\textwidth}
%     \centering
%     \includegraphics[width=\textwidth]{plots/rubrene/rubrene-holstein-imag-memory-temp-0.4to400K-freq-0to10omega-COLOUR.pdf}
%   \end{subfigure}
%   \caption{Frequency dependence of the polaron memory function from the variational parabolic Holstein and Fr\"ohlich models for a bulk 3D Rubrene organic crystal, for temperatures $T = 0.48$~\si{K}, $100$~\si{K}, $200$~\si{K}, $300$~\si{K} and $400$~\si{K}. The top row shows the real (left) and imaginary (right) components of the Fr\"ohlich memory function. The bottom row shows the real (left) and imaginary (right) components of the parabolic Holstein memory function.}
%   \label{fig:rubrene_memory}
% \end{figure*}

\begin{figure*}[!tbp]
    \centering
  \begin{subfigure}[b]{0.49\textwidth}
    \centering
    \includegraphics[width=\textwidth]{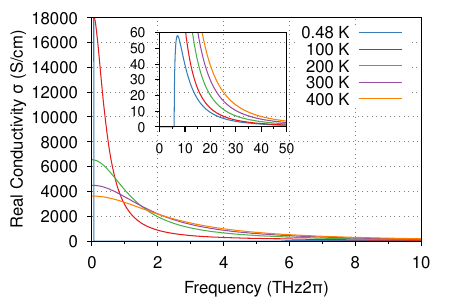}
  \end{subfigure}
  \hfill
  \begin{subfigure}[b]{0.49\textwidth}
    \centering
    \includegraphics[width=\textwidth]{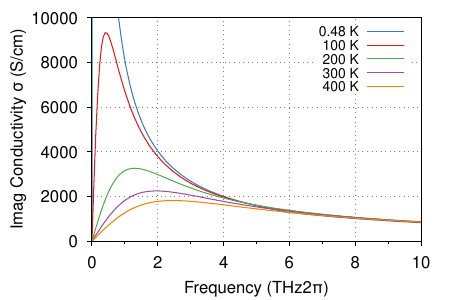}
  \end{subfigure}
  \begin{subfigure}[b]{0.49\textwidth}
    \centering
    \includegraphics[width=\textwidth]{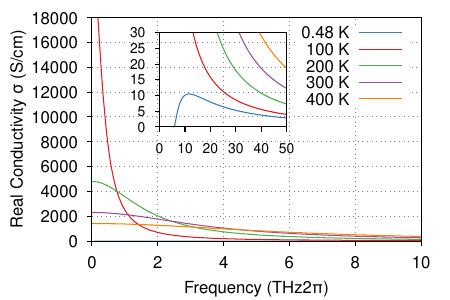}
  \end{subfigure}
  \hfill
  \begin{subfigure}[b]{0.49\textwidth}
    \centering
    \includegraphics[width=\textwidth]{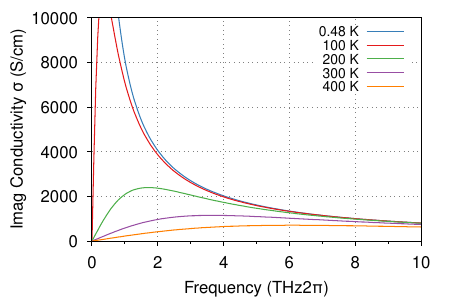}
  \end{subfigure}
  \caption{Frequency dependence of the polaron complex conductivity (in units of \si{\siemens\per\centi\meter}) predicted by the variational parabolic Holstein and Fr\"ohlich models for a bulk 3D Rubrene organic crystal, for temperatures $T = 0.48$~\si{K}, $100$~\si{K}, $200$~\si{K}, $300$~\si{K} and $400$~\si{K}. The top row shows the real (left) and imaginary (right) components of the Fr\"ohlich complex conductivity. The bottom row shows the real (left) and imaginary (right) components of the parabolic Holstein complex conductivity.}
  \label{fig:rubrene_conductivity}
\end{figure*}

In organic electronic materials, it is understood that the charge-carrier state is a small polaron.  This is often modelled with semi-classical transfer rate theories as a classical object hopping from site to site. The matrix elements which parameterise these rate equations can be calculated, within certain approximations, from electronic-structure calculations, but it is a challenge (and often input to the simulation and calculations) to define the sites on which the charge carriers are localised. 

One of the prototypical materials studied frequently to investigate electron-phonon coupling is Rubrene (5,6,11,12-tetraphenyltetracene) which has one of the highest carrier mobilities and can reach a few tens of cm$^2$/Vs for holes. This serves as a good test for applying our newly derived variational Holstein model for predicting its charge-carrier mobility in bulk. We take parameters for Rubrene from Ordejon et al.~\cite{Ordejn2017} where they derived Peierls (off-site) and Holstein (on-site) contributions by fitting the generalise Holstein-Peirels model with Density Functional Theory (DFT) calculations performed using SEISTA code. Here we make use of their Holstein parameters coupling, which we list in Table~\ref{tab:rubrene_data}, and use these parameters within our newly derived variational Holstein method. For simplicity, we consider a single effective phonon frequency, though the method presented here could be extended to multiple phonon modes, as we have demonstrated for the Fr\"ohlich Hamiltonian~\cite{MartinMultiple2022}.

The results for ground-state Holstein and Fr\"ohlich polarons for Rubrene are shown in Table~\ref{tab:gs_rubrene}. Likewise, Table~\ref{tab:rt_rubrene} gives the result for $T = 300$ K including the finite temperature DC mobility, which we calculate to be $\mu^{(H)}_{\text{Rubrene}} = 47.72$~\si{cm^2V^{-1}s^{-1}} for the Holstein polaron and $\mu^{(F)}_{\text{Rubrene}} = 92.59$~\si{cm^2V^{-1}s^{-1}} for the Fr\"ohlich polaron. Immediately, the Holstein prediction is more inline which whats observed in experiments whereas the Fr\"ohlich overestimates.

% In Fig.~\ref{fig:rubrene_temp} we present the temperature-dependent properties for the Rubrene polaron: polaron free energy and mobility. We provide the optimal variational parameters $v$ and $w$ with respect to temperature. In the top-left figure, we have the polaron \emph{binding} energies (relative to the band extrema) for Rubrene predicted by both models. The Holstein polaron binding energy is significantly smaller than the Fr\"ohlich polaron since it only ever couples to one lattice site, whereas the Fr\"olich polaron (in principle) couples to many lattice sites over an extent of multiple lattice constants. In the top-right figure, we have the temperature dependence of the $v$ and $w$ variational parameters. For the Fr\"ohlich polaron these take a minimum at $T=50$~\si{K} and then increase linearly above this temperature. For the Holstein polaron, the variational parameters have less temperature dependence, but the $w$ parameter also shows a minimum at $T=50$~\si{K} whereas the $v$ parameter is minimum towards $T=0$~\si{K}. In the bottom figure, we have the temperature dependence of the polaron charge-carrier mobility. The Holstein polaron mobility descends far more quickly than for the Fr\"ohlich polaron and reaches what will eventually become a constant value around $\mu \sim 10.0$ cm$^2$V$^{-1}$s$^{-1}$ towards higher temperatures, whereas the Fr\"ohlich mobility will continue to decrease at a rate proportional to $\mu \sim T^{-1/2}$.

\subsubsection{Rubrene Thermal Properties}

In Fig.~\ref{fig:rubrene_temp} we analyse the polaron properties as predicted by the variational parabolic Holstein and Fröhlich models in a bulk 3D Rubrene organic crystal. This figure includes three subplots: the polaron binding energy, the optimal variational parameters, and the DC polaron mobility, each as a function of temperature.

The top-left subplot of Fig.~\ref{fig:rubrene_temp} depicts the polaron binding energy, $F$, as a function of temperature ($T$). The binding energy is presented in millielectronvolts (meV). Both the Holstein (H) and Fr\"ohlich (F) models show a decrease in binding energy with increasing temperature. Notably, the Holstein model (blue curve) predicts a more pronounced decline in binding energy compared to the Fr\"ohlich model (red curve). At lower temperatures, the binding energy in the Holstein model is significantly higher, indicating stronger polaron formation. As temperature increases, the binding energies of both models converge, reflecting a reduction in polaron stabilisation due to thermal agitation.

The top-right subplot illustrates the optimal variational parameters $v$ (dimensionless) and $w$ (in units of $\omega_0$) as functions of temperature. The parameters $v_H$ and $v_F$ (blue and red curves, respectively) represent the Holstein and Fr\"ohlich models. Similarly, $w_H$ and $w_F$ (green and magenta curves, respectively) correspond to the $w$ parameters. Both variational parameters exhibit non-monotonic behaviour with temperature, with distinct minima observed around 120 K for the Holstein model and 50 K for the Fr\"ohlich mode. This suggests that at certain temperatures, the polaron wavefunction undergoes significant changes, potentially due to the competing effects of electron-phonon coupling and thermal excitation.

The bottom subplot shows the DC polaron mobility, $\mu(T)$, in units of $\text{cm}^2\text{V}^{-1}\text{s}^{-1}$, as a function of temperature. The mobility decreases sharply with increasing temperature for both models, indicating enhanced phonon scattering at higher temperatures. The two models predict similar mobilities below 120 K, with the Holstein mobility (blue curve) a little higher than the Fr\"ohlich model (red curve). However, the Fr\"ohlich model predicts consistently higher mobility compared to the Holstein model above this temperature. This can be attributed to the long-range nature of the electron-phonon interactions in the Fr\"ohlich model, which tends to preserve higher mobility despite thermal disruptions.

These results provide insights into the temperature dependence of polaron properties in 3D Rubrene crystals. The distinct behaviours of the Holstein and Fr\"ohlich models highlight the importance of considering different electron-phonon interaction mechanisms when analysing organic semiconductors. The findings underscore the complex interplay between thermal effects and electron-phonon coupling, which governs the formation, stability, and mobility of polarons in these materials.

\subsubsection{Rubrene Complex Conductivity}

% In Fig.~\ref{fig:rubrene_conductivity} we show the frequency-dependence of the real and imaginary components of the complex conductivity for both polarons. At low temperatures both polarons see a response peak beginning at the phonon frequency $\omega_0 = 5.768$ THz2$\pi$, but then the Fr\"ohlich polaron response decays far more rapidly with frequency than the Holstein polaron. As we increase the temperature, this trend continues, except that for both polarons we begin to see some response below the phonon frequency due to the presence of thermally excited phonons that generate an extra background response. This results in a local minimum in the conductivity around the effective polaron frequency $v$ as energy is lost to internal phonons that make up the polaron state. At much higher temperatures, the thermally excited phonon response now drowns out any kind of polaronic response and we are left with a typical Drude-like conductivity for both polarons, again with the Holstein polaron decaying more slowly than the Fr\"ohlich polaron with increasing frequency.

% By applying both polaron models to Rubrene, we can more clearly see that the physics described by either model is very different. However, the predictions of the Holstein model seem to better align with the experimentally observed charge-carrier mobility.

Fig.~\ref{fig:rubrene_conductivity} illustrates the frequency dependence of the polaron complex conductivity, presented in units of Siemens per centimetre ($\text{S}/\text{cm}$), as predicted by the variational parabolic Holstein and Fr\"ohlich models for a bulk 3D Rubrene organic crystal at various temperatures. This figure contains four subplots: the real and imaginary components of the Fr\"ohlich and Holstein complex conductivities, each plotted against frequency.

The top-left subplot of Fig.~\ref{fig:rubrene_conductivity} displays the real component of the Fr\"ohlich model's conductivity across a range of temperatures: 0.48 K, 100 K, 200 K, 300 K, and 400 K. At the lowest temperature (0.48 K), the real conductivity shows a significant one-phonon peak at low frequencies starting at the phonon frequency $\Omega = \omega_0 = 5.768$~\si{THz}, which rapidly decreases with increasing frequency. As temperature increases, the peak becomes less pronounced and is no longer visible, indicating a reduction in the polaronic response due to the presence of thermally excited phonons that generate an extra background response. The inset in the top-left subplot provides a closer view of the high-frequency behaviour, emphasising the one-phonon peak and the gradual decline in real conductivity across all temperatures.

The top-right subplot represents the imaginary component of the Fr\"ohlich model's conductivity for the same temperature range. Similar to the real component, the imaginary conductivity exhibits a peak at low frequencies, with the magnitude decreasing as the temperature rises. This peak indicates the reactive part of the polaron response, which is dominant at lower frequencies and diminishes with increasing thermal agitation.

The bottom-left subplot shows the real component of the Holstein model's conductivity at the specified temperatures. The conductivity profile is similar to that of the Fr\"ohlich model, with a prominent one-phonon peak at low frequencies starting at the phonon frequency that diminishes with increasing temperature to be replaced by a Drude-like background response. The inset highlights the high-frequency region, revealing the temperature-dependent decrease in real conductivity.

The bottom-right subplot illustrates the imaginary component of the Holstein model's conductivity. This component also shows a peak at low frequencies, which decreases in magnitude as temperature increases. The behaviour is consistent with the Fr\"ohlich model, although the absolute values and specific trends differ due to the distinct nature of electron-phonon interactions in the Holstein model.

These plots collectively demonstrate the frequency-dependent behaviour of polaron complex conductivity in 3D Rubrene crystals as predicted by the Holstein and Fr\"ohlich models. The observed trends underscore the significant impact of temperature on both the real and imaginary components of conductivity. At low temperatures, the strong polaronic interactions result in a high conductivity one-phonon peak at low frequencies, whereas increased thermal energy at higher temperatures disrupts these interactions due to thermally excited phonons, leading to a diminished polaronic response. These insights are crucial for understanding the dynamic electrical properties of organic semiconductors under varying thermal conditions.

\section{Discussion and Conclusion}

The study successfully extends the Feynman variational method to a Holstein small polaron with a parabolic band, allowing for a unified approach to predicting polaron mobility in organic semiconductors. This extension retains the same quasiparticle Lagrangian as the original 1955 theory~\cite{Feynman1955}, facilitating the use of FHIP response theory~\cite{Feynman1962} to calculate DC mobility and complex conductivity. The application of this theory to crystalline Rubrene demonstrates its validity, showing good agreement with experimental measurements of mobility.

One of the key findings is the identification of a discrete localisation transition as a function of coupling strength in the Holstein model. This transition is crucial for understanding charge transport in organic semiconductors, where the electron-phonon coupling is relatively strong. The theory's ability to take matrix elements from electronic structure calculations on real materials marks a significant advancement, as demonstrated by the accurate prediction of Rubrene mobility.

This research represents a significant advancement in the theoretical modelling of polaron mobility in organic semiconductors. By extending the Feynman variational method to the Holstein model, the study provides a robust framework that can predict the behaviour of polarons from weak to strong coupling regimes. The accurate comparison to diagrammatic Monte Carlo data validates the theory's successful application to Rubrene and opens the door for its use in other organic electronic materials.

Our analysis further highlights the theoretical and practical significance of extending the FHIP mobility theory to incorporate new variational solutions for different forms of electron-phonon coupling. Despite the age of the FHIP mobility theory, its adaptability and predictive power make it a valuable tool for modelling and designing novel semiconductors. The results of this study not only validate the theory's applicability but also emphasise the need for further research to explore its predictive capabilities across various applications, including intrachain mobilities in organic semiconductor polymer backbones.  

\section{Acknowledgement}
We are grateful for the discussions with Stefano Ragni and Thomas Hahn. 
The correct development of the variational methods was enabled by comparison to (unpublished) DiagMC data of Stefano Ragni, and we are grateful for additional `parabolic' DiagMC simulations they created for co-plotting in this paper. 
B.A.A.M. is supported by an EPSRC Doctoral Training Award. 
% TODO - do we have a code for the DTA ?
%
J.M.F. is supported by a Royal Society University Research Fellowship
(URF-R1-191292). 
%
%This work used the Imperial College Research Computing Service~\cite{HPC}.  
%Via our membership of the UK's HEC Materials Chemistry Consortium, which is
%funded by EPSRC (EP/R029431), this work used the ARCHER2 UK National
%Supercomputing Service (http://www.archer2.ac.uk).
% RSE?
%Julia\cite{Julia} codes implementing these calculations are available as a repository on
%GitHub\cite{GitHub}. 

\section*{Author contributions}
% See https://casrai.org/credit/ 
% Conceptualization
% Data curation
% Formal Analysis
% Funding acquisition
% Investigation
% Methodology
% Project administration
% Resources
% Software
% Supervision
% Validation
% Visualization
% Writing – original draft
% Writing – review & editing
The author contributions are defined with the Contributor Roles Taxonomy (CRediT). 
B.A.A.M.: Investigation, Formal analysis, Methodology, Software, Visualization, Writing – original draft. 
J.M.F.: Conceptualization, Investigation, Methodology, Software, Supervision, Writing – original draft.

\section*{Data access statement}

Codes implementing this new Holstein theory are present in our \textsc{PolaronMobility.jl} open-source package\cite{FrostJOSS2018}, along with a script that reproduces the calculations and generates the plots presented here. 

\bibliography{ProjectRubyPolarons}

\end{document}